# THE PRIVACY PARADOX IS A MISNOMER: DATA UNDER STRUCTURAL UNCERTAINTY

Ignacio Cofone*

*Abstract*

*The infamous privacy paradox refers to the apparent inconsistency between people's stated concern for privacy and their readiness to disclose personal information. This phenomenon has sparked two largely disconnected literatures: one offering experimental evidence of inconsistent behavior, and another providing qualitative accounts and defending the importance of privacy.*

*The Article presents an online field experiment that bridges those literatures and shows that the so-called paradox arises from a mischaracterization of the underlying behavior. The Article finds that it is structural uncertainty about risk that drives seemingly paradoxical privacy decisions. It does so by isolating discounting mechanisms and empirically testing whether observed privacy choices reflect temptation or rational responses to uncertainty. The results suggest that privacy behavior is not paradoxical but, rather, consistent with choices shaped by incomplete information.*

*The Article then discusses the policy implications of this reframing. As privacy decisions stem from structural uncertainty, which operates as a market failure, regulation should aim to reduce that uncertainty. This supports regulation that prioritizes transparency—for people to assess the risks of data collection—and flexibility mechanisms that accommodate evolving contexts. Such*

* Professor of Law & Regulation of AI, University of Oxford, Faculty of Law & Institute for Ethics in AI; Affiliated Fellow, Yale Information Society Project. Many thanks to Alessandro Acquisti, Ian Ayres, Jane Bambauer, Jack Balkin, Andrew Hayashi, Al Klevorick, Florencia Marotta-Wurgler, Cherie Metcalf, Alan Miller, Adriana Robertson, Ira Rubenstein, Robert Spear, Katherine Strandburg, and Tom Tyler for their helpful comments on prior versions of this Article. I'm also grateful to participants of the Canadian Law & Economics Association Conference, two anonymous reviewers contacted by the GLTR, and the editors of the GLTR for their helpful feedback; and I'm grateful to Aya Amer for her extraordinary research assistance. The Article was possible thanks to funding from the Yale Law School Oscar M. Ruebhausen Fund and the Social Sciences and Humanities Research Council of Canada Insight Development Grant.

*reframing provides a new argument for the right to be forgotten, which allows people to revisit prior disclosures as new risks become apparent. By shifting the focus from individual inconsistency to structural uncertainty, the findings call for privacy law to better reflect the reality of people's decision-making environments.*

**Keywords:** decision-making under risk, decision-making under uncertainty, hyperbolic discounting, privacy paradox, information privacy, nudging, transparency, privacy policies, right to be forgotten.

## TABLE OF CONTENTS

## INTRODUCTION

Why do people care about privacy but give it away for free? This Article argues that they do not; it just looks that way.

As we use (free or paid) digital products and services, the real cost is our personal data: Under the policies that govern apps, websites, and connected devices, their users gain access to a product or service, while agreeing to let providers collect, use, and monetize their personal information. In the process, the provider harvests data that ranges from banal (likes, browsing history) to revealing (relationships, locations, political views, and consumption patterns). This arrangement has led many to mistakenly assume that people engage in a calculus: that they disclose just enough information to balance the perceived benefit of using the service against the privacy risks they incur.[1]

Yet mounting evidence challenges that assumption. Consumers and advocacy groups express enormous dissatisfaction with the state of privacy in the information economy, complaining that their privacy is not properly protected.[2] Public opinion surveys consistently show that people value their privacy highly.[3]

---

[1] *See, e.g.*, Julien Cloarec, Lars Meyer-Waarden & Andreas Munzel, *Transformative Privacy Calculus: Conceptualizing the Personalization-Privacy Paradox on Social Media,* 41 PSYCH. & MKTG. 1574, 1574-75 (2024), https://doi.org/10.1002/mar.21998; Yedi Wang, Jiaji Zhu, Renhuai Liu & Yushi Jiang, *Enhancing Recommendation Acceptance: Resolving the Personalization-Privacy Paradox in Recommender Systems: A Privacy Calculus Perspective*, 76 INT'L J. INFO. MGMT., June 2024, at 1, 1–2, https://doi.org/10.1016/j.ijinfomgt.2024.102755; Jialin Fu, Jiaming Zhang & Xihang Li, *How Do Risks and Benefits Affect User' Privacy Decisions? An Event-Related Potential Study on Privacy Calculus Process*, FRONT. PSYCH., Feb. 16, 2023, at 1, 7, https://doi.org/10.3389/fpsyg.2023.1052782; *see also* Richard A. Posner, *The Right of Privacy*, 12 GA. L. REV. 393, 400 (1978) (discussing information disclosure tendencies under a rational choice model); George J. Stigler, *An Introduction to Privacy in Economics and Politics*, 9 J. LEGAL STUD. 623, 627 (1980) ("[I]n voluntary transactions there is no reason to interfere to protect one party provided the usual conditions of competition prevail; the efficient amount of information will be provided in transactions, given the tastes of the parties for knowledge and privacy."), https://doi.org/10.1086/467657; Jack Hirshleifer, *Privacy. Its Origin, Function, and Future*, 9 J. LEGAL STUD. 649, 662–64 (1980), https://doi.org/10.1086/467659.

In a number of incentivized experiments, the same people who express valuing their privacy disclose personal data for surprisingly low compensation.[4] The puzzling behavior of valuing privacy in theory but relinquishing it in practice came to be known as the "privacy paradox."[5]

A widespread explanation of the privacy paradox in experimental literature is that cognitive biases, particularly present bias (overweighting immediate costs and benefits relative to future ones),[6] lead people to inconsistent privacy choices in which, due to

---

[2] Pew Research Center, *Americans and Privacy: Concerned, Confused and Feeling Lack of Control over Their Personal Information* (2019) (indicating 81% of respondents believe the risks outweigh the benefits), https://www.pewresearch.org/internet/2019/11/15/americans-and-privacy-concerned-confused-and-feeling-lack-of-control-over-their-personal-information/ [https://perma.cc/SD9G-MH3E]; Patricia A. Norberg, Daniel R. Horne & David A. Horne, *The Privacy Paradox: Personal Information Disclosure Intentions versus Behaviors*, 41 J. CONSUMER AFFS. 100, 119 (2007) ("[T]he current trajectory is certainly for more versus less collection and use of personal information with consumers increasingly feeling like they have 'no place to hide'."), https://doi.org/10.1111/j.1745-6606.2006.00070.x.

[3] *See, e.g.*, Mallory Newall & Johnny Sawyer, *A Majority of Americans Are Concerned about the Safety and Privacy of Their Personal Data*, IPSOS (May 5, 2022), https://www.ipsos.com/en-us/news-polls/majority-americans-are-concerned-about-safety-and-privacy-their-personal-data [https://perma.cc/39GR-WSNC]; Colleen McClain, Michelle Faverio, Monica Anderson & Eugenie Park, *How Americans View Data Privacy*, Pew Research Center (Oct. 18, 2023), https://www.pewresearch.org/internet/2023/10/18/how-americans-view-data-privacy/ [https://perma.cc/A33U-64KK]; *see also* Office of the Privacy Commissioner of Canada, *2024–2025 Public Opinion Research on Privacy Issues*, https://www.priv.gc.ca/en/opc-actions-and-decisions/research/explore-privacy-research/2025/por_ca_2024-25/ [https://perma.cc/Q8S8-ZMRJ].

[4] *See infra* notes 10–14.

[5] *See* Nina Gerber, Paul Gerber & Melanie Volkamer, *Explaining the Privacy Paradox: A Systematic Review of Literature Investigating Privacy Attitude and Behavior*, 77 COMPUTS. & SEC. 226, 227 (2018) (providing a definition of the term), https://doi.org/10.1016/j.cose.2018.04.002; *see also* Byoungsoo Kim & Daekil Kim, *Understanding the Key Antecedents of Users' Disclosing Behaviors on Social Networking Sites: The Privacy Paradox*, 12 SUSTAINABILITY 5163, 5163-66 (2020), https://doi.org/10.3390/su12125163.

[6] *See, e.g.,* Alessandro Acquisti, Leslie John & George Loewenstein, *What is Privacy Worth?*, 42 J. LEGAL STUD. 249, 257 (2013) (providing a

their biases, they hyperbolically discount the long-term benefits of privacy protection.[7] This Article offers an alternative account. Previous research has also noted that information asymmetries affect privacy decision-making.[8] Building on those parallel insights, this Article provides an articulation of how a particular lack of information, structural uncertainty about future harms, drives privacy behavior.

This Article does so by presenting the first experiment that isolates the discounting mechanism at play for privacy decisions. The mechanism is isolated by testing preferences for pre-commitment vs flexibility in information disclosure decisions. Its findings indicate that the apparent inconsistency in behavior arises primarily from discounting under conditions of uncertain risk.[9] When people are unsure about what practices their information is subjected to, and the consequences of those data practices, they may disclose information while still valuing privacy because the perceived risk at the time of data collection is too uncertain to act upon decisively. The so-called privacy paradox, as a result, is not a paradox at all, but behavior that reflects adaptation to structural informational constraints.

Showing that privacy behavior is not paradoxical matters both conceptually and practically. If people are not reversing their

---

clear explanation of the theory), https://doi.org/10.1086/671754; Azim Shariff, Joe Green, & William Jettinghoff, *The Privacy Mismatch: Evolved Intuitions in a Digital World*, 30 CURR. DIR. PSYCH. SCI. 159, 159–64 (2021) (relating it to evolutionary theory), https://doi.org/10.1177/0963721421990355; Spyros Kokolakis, *Privacy Attitudes and Privacy Behaviour: A Review of Current Research on the Privacy Paradox Phenomenon*, 64 COMPUTS. & SEC. 122, 123 (2017), https://doi.org/10.1016/j.cose.2015.07.002; *see also* Gerber et al., *supra* note 5, at 229–30; Kim & Kim, *supra* note 5, at 2–5.

[7] Alessandro Acquisti, *Privacy in Electronic Commerce and the Economics of Immediate Gratification*, *in* PROCEEDINGS OF THE FIFTH ACM CONFERENCE ON ELECTRONIC COMMERCE 21, 24 (Jack Breese, Joan Feigenbaum & Margo Seltzer eds., 2004), https://doi.org/10.1145/988772.988777; *see also infra* note 43 (defining hyperbolic discounting).

[8] Daniel J. Solove, *The Myth of the Privacy Paradox*, 89 GEO. WASH. L. REV. 1, 4–5 (2021) (arguing that the privacy paradox is an illusion arising from a failure to distinguish between decisions and general attitudes), https://doi.org/10.2139/ssrn.3536265; Filippo Lancieri, *Narrowing Data Protection's Enforcement Gap*, 74 ME. L. REV. 15, 29–32 (2022); *see also* Alessandro Acquisti, Curtis Taylor & Liad Wagman, *The Economics of Privacy*, 54 J. ECON. LITERATURE 442, 445–46 (2016) (reviewing empirical evidence on the role of bounded rationality and information asymmetries in privacy behavior), https://doi.org/10.1257/jel.54.2.442.

[9] *See infra* note 43 and accompanying text.

preferences or acting irrationally, but instead responding to structural uncertainty, their behavior should not be seen as a failure of decision-making but as a failure of the information environment.

This distinction is crucial: When the problem is caused by temptation, regulation should intervene to correct individual behavior (for example, through nudges or default rules). But when the problem is caused by uncertainty about risk (i.e., about what data is collected, how it will be used, who it will be shared with, and with what consequences), the appropriate regulatory response is to develop rules that reduce uncertainty, for example, by improving transparency and allowing people to revise decisions as circumstances evolve. In short, diagnosing the correct mechanism behind the privacy paradox determines not only how we interpret behavior, but also how we design privacy law to address it.

The results of this experiment support a structural uncertainty account of the behavior that the privacy paradox literature observes, where people lack knowledge about underlying processes that are necessary to make the decisions they are asked to make. These results therefore provide arguments in favor of implementing targeted transparency obligations over data practices, more accessible privacy policies, and the right to be forgotten, which allows people to change their mind over their personal data.

The Article proceeds as follows. Part I reviews the empirical findings that define the privacy paradox. Part II explains the two different mechanisms, biased-based discounting and uncertainty-based discounting, that can explain the observed behavior. Part III presents an original experiment on information disclosure designed to test which mechanism accounts for the behavior. Part IV assesses the robustness of both accounts in light of the findings. Part V examines the regulatory implications of these findings with a particular focus on transparency and flexibility mechanisms such as the right to be forgotten.

## I. THE PRIVACY PARADOX

The "privacy paradox" is the supposed discrepancy between people's stated concern for privacy and their behavior.[10] The paradox

---

[10] Ruwan Bandara, Mario Fernando & Shahriar Akter, *Explicating the Privacy Paradox: A Qualitative Inquiry of Online Shopping Consumers*, 52 J. RETAILING & CONSUMER SERV. 1, 5–6 (2020), https://doi.org/10.1016/j.jretconser.2019.101947; *see also* Allison Woodruff, Vasyl Pihur, Sunny Consolvo, Lauren Schmidt, Laura Brandimarte & Alessandro Acquisti, *Would a Privacy Fundamentalist Sell Their DNA for $1000 . . . If Nothing Bad Happened as a Result? The Westin Categories, Behavioral Intentions, and Consequences*, *in* PROCEEDINGS OF

reflects that, in surveys, people express strong preferences for privacy but, in practice, they often disclose personal information for negligible rewards. This gap has led many to conclude that people do not truly value privacy. This Article argues that, instead, this gap reflects the impossibility of making decisions under structural uncertainty about how personal data will be used and what harm that entails.

### A. How People Value Privacy

How much do people actually value their privacy? Early empirical studies suggest a misalignment between people's declared concern for privacy and their actual online behavior.[11] Several studies grouped participants according to their declared level of privacy concern (high or low) and found that, during online shopping simulations, both groups disclosed the same amount of personal information.[12]

Many of these studies present the framing that privacy concerns announced prior to the experiment are inconsistent with shopping behavior during the experiment.[13] For example, subjects' privacy concerns, they explain, turn out to be a weak predictor of whether someone joins a social network and of how much information they share in it.[14] In one experiment, nearly 90% of respondents said they

---

THE 10TH SYMPOSIUM ON USABLE PRIVACY AND SECURITY 1, 1 (Lorrie Faith Cranor, Lujo Bauer & Robert Biddle eds., 2014), https://doi.org/10.1184/R1/6472181; Gerber et al., *supra* note 5, at 227; Kim & Kim, *supra* note 5, at 2.

[11] Sarah Spiekermann, Jens Grossklags & Bettina Berendt, *E-privacy in 2nd Generation E-Commerce: Privacy Preferences Versus Actual Behavior*, *in* PROCEEDINGS OF THE THIRD ACM CONFERENCE ON ELECTRONIC COMMERCE 38, 38–47 (Michael P. Wellman & Yoav Shoham eds., 2001); *see also* Susanne Barth & Menno D.T. de Jong, *The Privacy Paradox – Investigating Discrepancies Between Expressed Privacy Concerns and Actual Online Behavior – A Systematic Literature Review*, 34 TELEMATICS & INFORMATICS 1038 (2017), https://doi.org/10.1016/j.tele.2017.04.013.

[12] *See* Spiekermann et al. *supra* note 11, at 40, 44–45 (including information such as: in which occasions the subject takes photos, what she does with her pictures, what is her motivation for taking pictures, how photogenic she is, and how conceited she is).

[13] Bettina Berendt, Oliver Günther & Sarah Spiekermann, *Privacy in E-Commerce: Stated Preferences vs. Actual Behavior*, 48 COMMC'NS ACM 101, 104–05 (2005).

[14] Alessandro Acquisti & Ralph Gross, *Imagined Communities: Awareness, Information Sharing, and Privacy on the Facebook*, *in* PROCEEDINGS OF THE 6TH INTERNATIONAL CONFERENCE ON PRIVACY ENHANCING TECHNOLOGIES 36, 56–57 (George Danezis & Philippe Golle eds., 2006).

were highly concerned about privacy, but almost 90% agreed to provide their full name and address in exchange for a loyalty card, even when the data might be publicly disclosed.[15]

Other experiments show that people are unwilling to pay to protect their personal information. When participants were presented with two identical stores that differ in the nature of information requested (one that requested sensitive information and another that asked for non-sensitive information), they tended to buy from the cheapest store, even if it required more data collection.[16] And when the prices were equal, participants showed no preference between the two options.[17]

Valuations also show a large gap between how much people are willing to pay to protect their data and how much they would need to be paid to give it up—a gap that is larger than for any other goods. In one experiment testing this gap, people's average willingness to accept money (WTA) in exchange for their information to become public was more than five times greater than the willingness to pay (WTP) to protect it from becoming public (a WTA:WTP ratio of 5.47).[18] This gap is nearly double the average ratio that researchers find for other goods (typically 2.92).[19] In a related survey, most participants under one treatment were unwilling to pay even one dollar to avoid behavioral advertising, while most under another treatment were unwilling to accept a dollar to allow it.[20]

However, a growing body of literature argues that the privacy paradox is false. Some scholars argue that user behavior is rather a manifestation of power imbalances and manipulation.[21] Notably, Daniel Solove argues that the paradox stems from a mischaracterization of privacy attitudes.[22] He critiques the reliance on behavioral experiments that fail to account for the context-specific nature of privacy decision-making.[23]

---

[15] Alessandro Acquisti & Jens Grossklags, *Privacy and Rationality in Individual Decision Making*, 3 IEEE SEC. & PRIV. 26, 29 (2005), https://doi.org/10.1109/MSP.2005.22.

[16] Alastair R. Beresford, Dorothea Kübler & Sören Preibusch, *Unwillingness to Pay for Privacy: A Field Experiment*, 117 ECON. LETTERS 25, 26 (2012), https://doi.org/10.1016/j.econlet.2012.04.077.

[17] *Id.* at 27.

[18] Acquisti et al., *supra* note 6, at 267–68.

[19] *Id.*

[20] Aleecia M. McDonald & Lorrie Faith Cranor, *Beliefs and Behaviors: Internet Users' Understanding of Behavioral Advertising*, 38 TELECOMM. POL'Y RSCH. CONF. 1, 25–26 (2010).

[21] Ari Ezra Waldman, *Cognitive Biases, Dark Patterns, and the 'Privacy Paradox'*, 31 CURR. OPIN. PSYCH. 105, 107 (2020), https://doi.org/10.1016/j.copsyc.2019.08.025; Neil Richards & Woodrow Hartzog, *The Pathologies of Digital Consent*, 96 WASH. U. L. REV. 1461 (2019).

### B. CONTEXT, SENSITIVITY, AND SALIENCE

People reveal quite different types of information: offline and online, sensitive and non-sensitive, etc. But many privacy paradox experiments treat all personal information as if it were interchangeable and equally meaningful (what economists call fungible). This can lead to inaccurate interpretations of experimental results because such treatment ignores the varying types of harm associated with different data practices and varying sensitivity associated with different kinds of data: privacy harms can range from minor annoyances, like spam email, to serious consequences, such as discrimination, depending on how the data is used and who is using it.[24]

People value different types of personal information differently.[25] For instance, people make more efforts to protect sensitive data than they do for non-sensitive data.[26] One empirical study found that, on average, people value their offline data (such as birth date) three times as much as their online data (such as their browsing history).[27] People also assign different values to their offline data depending on the type of information involved.[28] And values can vary within the same type of data. When asked to reveal their weight and age, for example, people asked for more money to reveal traits they perceived as undesirable, even if there were no direct consequences for doing so.[29] The more sensitive or private the trait, the higher the monetary value people place on it.

---

[22] Solove, *supra* note 8.

[23] *Id.*

[24] *See, e.g.*, Kirsten Martin & Helen Nissenbaum, *Measuring Privacy: An Empirical Test Using Context to Expose Confounding Variables*, 18 COLUM. SCI. & TECH. L. REV. 176, 180 (2016) (arguing that not specifying the context in survey questions about privacy makes questions so ambiguous that responses to them should not be considered informative); Ignacio Cofone, *A Healthy Amount of Privacy: Quantifying Privacy Concerns in Medicine*, 65 CLEV. ST. L. REV. 1, 6–7 (2017); Barth & De Jong, *supra* note 11, at 1052.

[25] *See, e.g.*, Tobias Dienlin & Sabine Trepte, *Is the Privacy Paradox a Relic of the Past? An In-Depth Analysis of Privacy Attitudes and Privacy Behaviors*, 45 EUR. J. SOC. PSYCH. 285, 289–95 (2015), https://doi.org/10.1002/ejsp.2049.

[26] *See* Martin & Nissenbaum, *supra* note 24 (considering contextual factors such as the type of information); Hui Na Chua, Jie Sheng Ooi & Anthony Herbland, *The Effects of Different Personal Data Categories on Information Privacy Concern and Disclosure*, 110 COMPUT. & SECUR., Aug. 2021, at 12, https://doi.org/10.1016/j.cose.2021.102453; Wenjing Xie & Kavita Karan, *Consumers' Privacy Concern and Privacy Protection on Social Network Sites in the Era of Big Data: Empirical Evidence from College Students*, 19 J. INTERACT. ADVERT. 187 (2019), https://doi.org/10.1080/15252019.2019.1651681; David L. Mothersbaugh, William K. Foxx II, Sharon E. Beatty & Sijun Wang, *Disclosure Antecedents in an Online Service Context: The Role of Sensitivity of Information,* 15 J. SERVICE RESEARCH 76, 91 (2012), https://doi.org/10.1177/1094670511424924.

People respond to security measures when they are visible. When privacy-related information is shown directly on search engines, people prefer websites that offer stronger privacy protections, especially when they are making purchases involving sensitive information.[30] When privacy policies are available and their content salient, people are willing to pay a premium to purchase from retailers that protect their privacy.[31] Salient design elements that are often called "visceral"—such as human-like features on websites, self-focused attention mechanisms, or formal visual web design—can improve understanding and also influence how much information people disclose.[32] Similarly, other studies show that people sometimes consciously weigh the potential benefits of sharing data (such as personalization and discounts) against the perceived risks (such as data misuse).[33] In sum, people's ability to make privacy decisions changes with complexity—there is not a general inability to deal with them.[34] This shows that people's decisions about privacy are often consistent within specific contexts, challenging the existence of a universal privacy paradox.[35]

---

[27] Juan Pablo Carrascal, Christopher Riederer, Vijay Erramilli, Mauro Cherubini & Rodrigo de Oliveira, *Your Browsing Behavior for a Big Mac: Economics of Personal Information Online*, *in* PROCEEDINGS OF THE 22ND INTERNATIONAL CONFERENCE ON WORLD WIDE WEB 189, 196 (Daniel Schwabe, Virgílio Almeida & Hartmut Glaser eds., 2013).

[28] Anya Skatova, Rebecca McDonald, Sinong Ma & Carsten Maple, *Unpacking Privacy: Valuation of Personal Data Protection*, PLOS ONE, May 2023, at 1; *see also* Helen Nissenbaum, *Contextual Integrity Up and Down the Data Food Chain* 20 THEORETICAL INQUIRIES L. 221, 230 (2019).

[29] Bernardo A. Huberman, Eytan Adar & Leslie R. Fine, *Valuating Privacy*, 3 IEEE SECUR. PRIV. MAG. 22, 22–25 (2005) (measuring desirability (self-perception) with post-experiment questionnaires).

[30] Julia Gideon, Lorrie Cranor, Serge Egelman & Alessandro Acquisti, *Power Strips, Prophylactics, and Privacy, Oh My!*, 2 SYMP. ON USABLE PRIV. & SEC. 133, 143 (2006); *see also* Kai-Lung Hui, Hock Hai Teo & Sang-Yong Tom Lee, *The Value of Privacy Assurance: An Exploratory Field Experiment*, 31 MIS Q. 19, 26–27 (2007), https://doi.org/10.2307/25148779.

[31] Janice Tsai, Serge Egelman, Lorrie Cranor & Alessandro Acquisti, *The Effect of Online Privacy Information on Purchasing Behavior: An Experimental Study*, 22 INFO. SYS. RSCH. 254, 255 (2011), http://doi.org/10.1287/isre.1090.0260.

[32] Victoria Groom & Ryan Calo, *Reversing the Privacy Paradox: An Experimental Study*, *in* 39 TELECOMMUNICATIONS POLICY RESEARCH CONFERENCE 1, 4 (2011); *see also* Nico Ebert, Kurt Alexander Ackermann & Björn Scheppler, *Bolder is Better: Raising User Awareness Through Salient and Concise Privacy Notices*, *in* PROCEEDINGS OF THE 2021 CHI

People's privacy concerns are not only about immediate harms, such as fear of fraud or spam; they are also about indirect consequences, such as price discrimination.[36] They are often more concerned about how their data will be used than about whether it will be shared.[37] Giving people control over the publication of their personal data lowers their privacy concerns and makes them more willing to share sensitive information.[38]

The findings of these studies suggest that user behavior might be less random than one might conclude from the privacy paradox literature.[39] When the context of data sharing is taken into account,

---

CONFERENCE ON HUMAN FACTORS COMPUTING SYSTEMS 1, 4–6 (2021) (providing evidence that bold, concise, and visually salient notices improve awareness and recall of privacy risks).

[33] Nina Gerber, Paul Gerber & Melanie Volkamer, *Explaining the Privacy Paradox: A Systematic Review of Literature Investigating Privacy Attitude and Behavior*, 77 COMPUT. & SEC. 226, 252 (2018), https://doi.org/10.1016/j.cose.2018.04.002; Emilie Storslett Henriksen, Asbjørn Følstad & Konstantinos Boletsis, *Exploring Users' Privacy Decision Making in Retail–Insights and Challenges for HCI Research*, QUALITY & USER EXPERIENCE, July 12, 2025, at 9–10, 15–16.

[34] Leslie John, Alessandro Acquisti & George Loewenstein, *Strangers on a Plane: Context-Dependent Willingness to Divulge Sensitive Information*, 37 J. CONSUMER RSCH. 858, 868 (2011), https://doi.org/10.1086/656423.

[35] *See* Solove, *supra* note 8, at 26–29; Helen Nissenbaum, PRIVACY IN CONTEXT: TECHNOLOGY, POLICY, AND THE INTEGRITY OF SOCIAL LIFE 127–48 (2010).

[36] Daniel J. Solove & Danielle Keats Citron, *Risk and Anxiety: A Theory of Data-Breach Harms*, 96 TEX. L. REV. 737, 777 (2018); M. Ryan Calo, *The Boundaries of Privacy Harm*, 86 IND. L.J. 1131, 1133 (2011); Luc Wathieu & Allan Friedman, *An Empirical Approach to Understanding Privacy Valuation* 8 (Harv. Bus. Sch. Working Paper No. 07-075, 2007); *see also* Andrew Odlyzko, *Privacy, Economics, and Price Discrimination on the Internet*, *in* ECONOMICS OF INFORMATION SECURITY 187, 188 (2003), https://doi.org/10.1007/1-4020-8090-5_15.

[37] WATHIEU & FRIEDMAN, supra note 36.

[38] *See* James A. Mourey & Ari Ezra Waldman, *Past the Privacy Paradox: The Importance of Privacy Changes as a Function of Control and Complexity*, 5 J. ASS'N CONSUMER RSCH. 162 (2020).

[39] *See supra* Section I.A.

people often behave quite rationally in making simple privacy choices.[40]

In light of these findings, some scholars have proposed bridging the gap between theoretical critiques of the privacy paradox and its experiments and implementations, advocating for new experimental approaches to test the effectiveness of different privacy interventions.[41]

The Parts that follow introduce such an approach. The two characterizations of privacy behavior (as inconsistent or as context-dependent) each imply different motivations for their decision-making. The next section outlines these motivations and how they can be experimentally distinguished.

## II. TWO DISCOUNTING MECHANISMS AND WHY THEY MATTER

Understanding why a person might not choose to protect their own privacy, even when they have an interest in doing so, requires understanding discounting. When people choose the less beneficial of two potential payoffs that will happen at different times, it is because they discount the value of the future one. There are two reasons to discount payoffs: the inconvenience of waiting and the risk of the payoff disappearing. A "payoff" in this sense can be positive or negative, and when it is negative it is called a penalty. Based on these discounting reasons, economists and psychologists identify a pattern of behavior called choice reversal: When one plans to avoid a big penalty in the future by taking a small penalty in the near future, but as the time to implement that plan gets closer, one chooses to avoid the small penalty.[42] An example is when one plans to clean cooking utensils directly after eating dinner, knowing it will be harder to clean them later on, yet avoids cleaning them when dinner is finished. The two possible reasons to discount payoffs and penalties (i.e., inconvenience and risk) create distinct causes for people to reverse choices, and knowing which one is at play matters for understanding and regulating behavior. The inconvenience of waiting, which leads to choice reversal when temptation exists, relies on assumptions about people making those decisions. The cost of risk, which can lead to choice reversal when it is uncertain, relies on assumptions about the decision-making context.

---

[40] Huberman, *supra* note 29 (finding that the likelihood of participants to disclose weight and age information varied depending on a trait's "desirability," highlighting the decision's "strongly contextual nature).

[41] Ida Adjerid, Eyal Peer & Alessandro Acquisti, *Beyond the Privacy Paradox*, 42 MIS Q. 465, 467, 472 (2018), https://doi.org/10.25300/MISQ/2018/14316; Barth & De Jong, *supra* note 11, at 1050–52.

[42] Or, conversely, when one plans to obtain a big payoff in the future by abandoning a small payoff now but, as the time to implement that time gets closer, one chooses to seize the small payoff.

### A. THE TEMPTATION ACCOUNT: DISCOUNTING BASED ON BIASES

One way to explain the privacy paradox is through an account in which people face temptation and end up overvaluing present outcomes when compared to future outcomes—a result of being present-biased in which people hyperbolically discount the future.[43] When people say they value their privacy highly but then disregard it, they are, in a way, setting a plan (to only give up their privacy for a high reward) and then deviating from it (by giving it up for a small reward).[44]

To many, online behavior seems consistent with findings in behavioral science that suggest that people often place less value on outcomes that are further in the future compared to those that are more immediate (i.e., they discount the distant future at a higher rate than the near future). Behavioral research has shown that people frequently choose immediate rewards over long-term benefits (and they face large negative consequences in the future to avoid a small immediate penalty) not because they genuinely prefer the short-term benefits but because of temptation (self-control) problems that affect their ability to make rational decisions.[45] Some argue that the privacy paradox is a case in which these biases drive behavior.[46] They explain behavior with an account according to which people

---

[43] Hyperbolic discounting is an increasing rate of time preference over time so that the distant future is discounted more heavily than the near future. *See* Christopher F. Chabris, David I. Laibson & Jonathon P. Schuldt, *Intertemporal Choice*, *in* 4 THE NEW PALGRAVE DICTIONARY ECON. 536, 536–42 (Steven N. Durlauf & Lawrence E. Blume eds., 2d ed. 2008).

[44] Richard Thaler, *Some Empirical Evidence on Dynamic Inconsistency*, 8 ECON. LETTERS 201, 205–206 (1981), https://doi.org/10.1016/0165-1765(81)90067-7.

[45] Ted O'Donoghue & Matthew Rabin, *Choice and Procrastination*, 116 Q.J. ECON. 121, 122–25, 148–49 (2001).

[46] *See, e.g.*, Acquisti, John & Loewenstein, *supra* note 6, at 257–58; *see also* Alessandro Acquisti & Jens Grossklags, *Privacy Attitudes and Privacy Behavior*, *in* ECONOMICS OF INFORMATION SECURITY 165 (L. Jean Camp & Stephen Lewis eds., 2004); Waldman, *supra* note 21, at 105; Kirsten McNally, '*Accept All'. How Hyperbolic Discounting Renders PSM a Faulty Foundation for Privacy Protection*, 3 STUD. PHIL. POLI. ECON. 1, 40 (2021).

downplay (i.e., discount) the seriousness of future privacy harms too much in favor of immediate rewards.[47]

The temptation account, in other words, interprets online behavior as driven by the tendency to disproportionately favor smaller, immediate rewards over larger, delayed ones. When people have a choice between accessing a digital service, such as a social media platform, right away by agreeing to share personal data, versus the delayed and less tangible benefit of avoiding future privacy harms, they tend to choose the former. The immediate reward comes from the instant usefulness of the service; the value of privacy, by contrast, is delayed and harder to picture. The temptation account builds on hyperbolic discounting literature, arguing people undervalue those delayed privacy benefits compared to the immediate convenience of the service, even if they believe privacy is important to them abstractly. What appears to be inconsistent behavior under the privacy calculus view is, under this account, predictable.

### B. THE UNCERTAINTY ACCOUNT: DISCOUNTING BASED ON RISK

Behavioral scientists have also pointed out that it is often unrealistic to assume that people can assign precise probabilities to future events or even that they have certain beliefs about what exactly those probabilities are.[48] In the information economy, people face uncertainty about what might happen when they share data: how serious the consequences might be, what steps they can take to protect themselves, what others are doing to safeguard their data, and what unexpected developments might occur.[49] This uncertainty plays a bigger role in online behavior than previously assumed. Work on behavioral science shows that uncertainty about risk can reconcile non-exponential discounting with consistent decision-making over time (i.e., dynamic consistency).[50] This theory offers an alternative account of people's online behavior when faced with privacy-related decisions.

---

[47] Acquisti & Grossklags, *Privacy Attitudes and Privacy Behavior, supra* note 46, at 129–30. *See generally* O'Donoghue & Rabin, *supra* note 45.

[48] Alessandro Acquisti & Jens Grossklags, *What Can Behavioral Economics Teach Us About Privacy?*, *in* DIGITAL PRIVACY: THEORY, TECHNOLOGIES AND PRACTICES 367(Alessandro Acquisti, Stefanos Gritzalis, Costos Lambrinoudakis & Sabrina De Capitani di Vimercati eds.,2007) ("[W]e favor the view that in numerous privacy-sensitive situations it is unrealistic to assume existence of known or knowable probabilities or complete (subjective) beliefs for probabilities over all possible outcomes."); Acquisti & Grossklags, *supra* note 15 (challenging earlier-held beliefs). *See generally* Alessandro Acquisti, Laura Brandimarte & George Loewenstein, *Privacy and Human Behavior in the Age of Information*, 347 SCIENCE 509 (2015) (documenting that users lack complete information), https://doi.org/10.1126/science.aaa1465.

Generally, people care less about future consequences than about present ones.[51] One reason for that is that waiting for things can be inconvenient. The other reason is that future events are uncertain: over time, benefits have a risk of depreciating or disappearing and we cannot know how bad costs might be. So accepting a benefit or incurring a cost now is (and feels) more concrete than accepting a benefit or incurring a cost for the future, which might or might not happen.

Imagine two scenarios. The first is one where an individual must choose between a payoff of $100 now (payoff *V* at time *T*) or a larger payoff of $150 in a year (payoff *V'* at later time *T'*). The second scenario is one in which the individual must choose between the same payoffs ($100 and $150) but they both have a delay: their timing is instead in three months or in a year and three months, respectively (*V* at *T+t* or *V'* at *T'+t*).

---

[49] In addition, these discounting models abstract from liquidity constraints, and therefore from immediate needs that a subject could have when facing the choice.

[50] *See generally* Thomas Epper, Helga Fehr-Duda & Adrian Bruhin, *Viewing the Future Through a Warped Lens: Why Uncertainty Generates Hyperbolic Discounting*, 43.3 J. RISK & UNCERTAINTY 169 (2011), http://doi.org/ 10.1007/s11166-011-9129-x; G. W. Story, Z. Kurth-Nelson, M. Moutoussis, K. Iigaya, G. J. Will, T. U. Hauser, B. Blain, I. Vlaev, and R. J. Dolan, *Discounting Future Reward in an Uncertain World*, 11 DECISION 255, 267-71 (2024), https://doi.org/10.1037/dec0000219. Traditional economic models assume exponential discounting, meaning that people apply a constant discount rate to future outcomes over time. But empirical studies show that people often discount the near future more steeply than the distant future, leading to choice reversals.

[51] Drazen Prelec & George Loewenstein, *Decision Making Over Time and Under Uncertainty: A Common Approach*, 37(7) MGMT. SCI. 770, 784 (1991), https://doi.org/10.1287/mnsc.37.7.770. *See generally* G. Ainslie, *Specious Reward:Behavioral Theory of Impulsiveness and Impulse Control*, 82(4) PSYCH. BULL. 463 (1975), https://doi.org/10.1037/h0076860.

| | **Earlier payoff** | **Later payoff** |
|---|---|---|
| **Scenario 1** | $100 now (*V*;*T*) | $150 in a year (*V'*;*T'*) |
| **Scenario 2** | $100 in 3 months (*V; T*+*t*) | $150 in 1 year and 3 months (*V'*; *T'*+*t*) |

Table 1: Illustrates the hypothetical scenario

Imagine that, in both scenarios, the promisor of the $150 amount has a small and stable chance of going bankrupt every year (i.e., the risk of losing the payoff is a linear function of time). If that is the case, a rational individual will discount the payoffs at a constant rate. When the individual has a choice between *V* at *T* and *V'* at *T'* with a linear risk (call it λ), the expected payoff to which they compare *V* should be $e^{-\lambda T'}V'$, which leads them to discount the value of the expected payoff by the same amount for every unit of time waited.[52] The same analysis applies to the second scenario.

But imagine that, on the other hand, the risk per period is not linear and the individual does not know the risk in each period (while there always remains some risk of the payoff disappearing).[53] For example, the promisor of the $150 has a small and *unstable* chance of going bankrupt every year (i.e., the risk is not a linear function of time) and the individual does not know how safely the promisor handles the business. Due to the uncertain risk, a rational individual will be more afraid of the payoff disappearing at the beginning of the waiting period.[54] Therefore, they will discount the payments in the first scenario ($100 now or $150 in a year) and the second scenario ($100 in three months or $150 in a year and three months) differently.[55] Their per-period discount for payoffs in the near future will be higher than their per-period discount payoffs for the more

[52] Peter D. Sozou, *On Hyperbolic Discounting and Uncertain Hazard Rates*, 265 PROC.: BIO. SCIS. 2015 (1998), https://doi.org/10.1098/rspb.1998.0534; Partha Dasgupta & Eric Maskin, *Uncertainty and Hyperbolic Discounting*, 95 AM. ECON. REV. 1290, 1291–92 (2005).

[53] i.e., $\lambda = \lambda'(T)$ and $\lambda' < 0$.

[54] Sozou, *supra* note 52.

[55] Epper et al., *supra* note 50, at 172–73.

distant future.[56] Since the individual is uncertain of the risk level in both scenarios, they will worry that the benefit could disappear in the period between $T$ and $T'$. As a result, they will apply a different discount rate at $T'$ than the one they used at $T$.[57]

This increased discount rate for more immediate payoffs leads to behavior that might look like the individual changed their mind.[58] Because they are applying the high discount now but, as time passes and they are closer to the later payoffs they will apply a lower discount,[59] the individual will choose \$100 ($V$) in the first scenario but \$150 ($V'$) in the second. This "choice reversal" is when one switches from wanting one option ($V$) to wanting a different one ($V'$) based on how far in the future the same alternatives are.

In other words, if the risk is not linear and unknown to the individual, a rational individual will still discount the future at a higher rate than the present.[60] Their choice in the first scenario matters if the benefit is available now (in technical terms, if the payoff "survives" until time $T$). But their choice in the second scenario only matters if the benefit is still available in three months (in technical terms, if the payoff survives until time $T+t$). In the first scenario, there is no risk associated with the present payoff, while the future payoff option is uncertain; but that is not true in the second scenario. They will appear to behave less patiently in one scenario than in the other, even if the actual risk were to stay the same.[61]

Now consider an example that is slightly different from the traditional economics setup above: instead of choosing between a small reward now and a larger one later, imagine someone choosing between a small cost now—such as spending time adjusting privacy settings or reading a long privacy policy—and the risk of a serious negative consequence later, such as a data breach or misuse of personal information.[62] In privacy decisions, the analogy to "payoff survival" flips: it is not about a reward disappearing, but about a harm materializing.[63] If the person believes the risk of future harm is vague or hard to visualize, they may downplay it. People may reason that if nothing bad has happened yet, perhaps nothing will, and treat the future risk as less pressing than the present inconvenience. Even if that harm could be serious, its uncertainty and delay will lead people to give it less weight than the immediate inconvenience.[64]

---

56 Dasgupta & Maskin, *supra* note 52, at 1292–94. The same will happen if the risk per period is declining.

57 Shane Frederick, George Loewenstein & Ted O'Donoghue, *Time Discounting and Time Preference: A Critical Review*, 40 J. ECON. LITERATURE 351, 361 (2002).

58 A discount rate reflects how much less a person values a future payoff compared to an immediate one. The higher the discount rate, the more heavily the future is devalued.

59 Frederick, Loewenstein & O'Donoghue, *supra* note 57.

60 Epper et al., *supra* note 50, at 187–92; *see also* Kota Saito, *A Relationship Between Risk and Time Preferences*, 101(5) AM. ECON. REV. 2271 (2011).

61 Sozou, *supra* note 52, at 2017; Yoram Halevy, *Time Consistency: Stationarity and Time Invariance*, 83 ECONOMETRICA 335, 348 (2015) [hereinafter Halevy, *Time Consistency*], https://doi/org/10.3982/ECTA10872; Omar Azfar, *Rationalizing Hyperbolic Discounting*, 38 J. ECON. BEHAV. ORG. 245, 248–251 (1999), https://doi.org/10.1016/S0167-2681(99)00009-8. *See generally* Yoram Halevy, *Strotz Meets Allais: Diminishing Impatience and the Certainty Effect*, 98 AM. ECON. REV. 1145–62 (2008) [hereinafter Halevy, *Strotz Meets Allais*].

This explains why people who care about their privacy might still take actions that expose them to long-term risks. In these cases, they are not trading a small benefit for a larger one, but rather avoiding a hassle now and, in doing so, exposing themselves to a possibly greater cost later. When the future harm is uncertain or abstract (i.e., its likelihood or severity is unclear), people tend to discount it heavily—meaning they give it much less weight in their decision-making than they would if it were certain. This kind of discounting is common in situations involving uncertain harms, like those linked to privacy.[65]

In sum, people who factor in unknown risk are increasingly likely to choose smaller, short-term rewards as the time to their first possible reward gets shorter (i.e., they show delay-dependent discounting), even when the time gap between their two potential rewards stays the same, while still behaving rationally.[66] This pattern holds regardless of whether the unknown risk remains, decreases, or increases.[67] In situations of uncertainty, non-expected utility models fit this kind of rational behavior.[68] So, a rational person facing equivalent choices (under uncertainty) can still show delay-dependent discounting and reverse their choices over time.[69] That means that, in contexts of uncertain risk, choices to avoid smaller short-term penalties despite a higher risk of larger long-term penalties do not necessarily imply a behavioral bias.[70] The so-called privacy paradox might therefore reflect a rational response to uncertainty about privacy harms: because people do not know how likely a future data breach or another privacy harm is,[71] they may understandably choose the immediate benefits involved in sharing their data.

---

[62] *See* Dan Ariely & Klaus Wertenbroch, *Procrastination, Deadlines, and Performance: Self-Control by Precommitment*, 13 PSYCH. SCI. 219, 222-23 (2002), https://doi.org/10.1111/1467-9280.00441.

[63] *See* Danielle K. Citron & Daniel J. Solove, *Privacy Harms*, 102 B.U. L. REV. 793, 816–22 (2022), http://dx.doi.org/10.2139/ssrn.3782222.

[64] Marianna Blackburn & Wael El-Deredy, *The Future is Risky: Discounting of Delayed and Uncertain Outcomes*, 94 BEHAVIOURAL PROCESSES 9 (2013), https://doi.org/10.1016/j.beproc.2012.11.005.

[65] *See* Daniel J. Solove, *The Limitations of Privacy Rights*, 98 NOTRE DAME L. REV. 975, 978 (2013) ("Although the benefits [of disclosure] are immediate and concrete, the costs involve risks that are more abstract and speculative.").

[66] Sozou, *supra* note 52, at 2016–17; Azfar, *supra* note 35, at 248–251.

[67] Halevy, *Strotz Meets Allais, supra* note 61, at 1156.

[68] Blackburn & Wael El-Deredy, *supra* note 64.

### C. HOW TO TELL TEMPTATION FROM UNCERTAINTY

It is possible to test for present-bias while controlling for uncertainty.[72] This can be done by presenting people with consumption choices involving immediate and delayed rewards, while introducing small changes before each decision. This approach helps isolate whether their behavior is driven by how they value time (temptation) or by the context (uncertainty about future outcomes).[73] Moving the time horizon forward, meaning that both options in a given decision are shifted further into the future, makes it so that neither choice involves an immediate gratification option that can trigger temptation; this is a standard method used to test for time inconsistency in economics.[74]

---

[69] This is also the case for uncertain delays, which involve fewer assumptions since the risk is certain but only its time of execution is uncertain. *See* Joseph T. McGuire & Joseph W. Kable, *Decision Makers Calibrate Behavioral Persistence on the Basis of Time-Interval Experience*, 124 COGNITION 216, 217–18 (2012), https://doi.org/10.1016/j.cognition.2012.03.008; Joseph T. McGuire & Joseph W. Kable, *Rational Temporal Predictions Can Underlie Apparent Failures to Delay Gratification*, 120 PSYCH. REV. 395–410 (2013).

[70] Halevy, *Time Consistency*, *supra* note 61, at 1145; Halevy, *Strotz Meets Allais*, *supra* note 61, at 348.

[71] Katherine J. Strandburg, *Free Fall: The Online Market's Consumer Preference Disconnect,* 430 U. CHICAGO LEG. F. 95, 130–31 (2013).

[72] Blackburn & El-Deredy, *supra* note 64.

[73] In technical terms, people receiving different changes (which the literature calls shocks) initially make different decisions regarding their consumption choices, but those become irrelevant as the time horizon is moved forward. *See* Jesus Fernandez-Villaverde & Arijit Mukherji, *Can We Really Observe Hyperbolic Discounting?* (Penn Inst. for Econ. Rsch., Working Paper No. 02-008, 2006). *See generally* GREGORY BESHAROV & BENTLEY COFFEY, RECONSIDERING THE EXPERIMENTAL EVIDENCE FOR QUASI-HYPERBOLIC DISCOUNTING DUKE DEPARTMENT OF ECONOMICS WORKING PAPER 1–22 (2003).

[74] Shifting both available options further into the future by the same amount of time removes the immediacy of the decision and allows

The way to discern between delay-dependent discounting caused by time preferences and delay-dependent discounting caused by uncertainty is to examine whether people prefer precommitment or flexibility.[75] Table 2, below, summarizes this contrast.

| | **Bias-based Discounting** | **Uncertainty-based Discounting** |
|---|---|---|
| **Mechanism** | Immediate gratification outweighs future risks | Decisions made under ambiguous risk due to incomplete information |
| **Assumed irrationality** | Yes: self-control failure (short-term biased) | No (operating under epistemic constraint) |
| **Structure** | Stable risk, varying privacy valuation | Unknown risk, stable privacy valuation |

Table 2: Types of discounting

A feature of temptation (self-control) problems is that people tend to recognize mismatches between their behavior and their long-term goals—they notice behavior that is inconsistent with their aims. They usually want to stop that behavior and, if they are what economists call "sophisticated," they also recognize that they are researchers to test whether participants' preferences change simply because one of the original options was available right away. If shocks influence behavior only when an immediate option is present, their effect should diminish once both options are future-oriented. In the initial setup, participants might choose between a smaller payoff (or lower cost) available now and a larger payoff (or higher cost avoided) available later. To move the time horizon forward, the experiment shifts both options into the future by the same amount. So now the participants are choosing between a smaller payoff in, say, 3 months and a larger payoff in, say, 6 months. In the field experiment described in Part III, the time horizon is moved by having rewards arrive later independently of the moment in which people make the choice. See Id.

[75] Marco Casari, *Pre-Commitment and Flexibility in a Time Decision Experiment*, 38 J. RISK & UNCERTAINTY 117, 118–19 (2009); Todd Rogers, Katherine L. Milkman & Kevin G. Volpp, *Commitment Devices: Using Initiatives to Change Behavior*, 311 JAMA 2065 (2014); *see also* Blackburn & El-Deredy, *supra* note 64, at 11–12 (using an alternative experimental design focused on uncertainty rather than on both uncertainty and temptation).

likely to continue that behavior unless they take steps to prevent it.[76] For these individuals, pre-committing to their goal becomes the best strategy.[77] People in this situation (facing temptation while aware of it) will value ways to bind themselves to their preferred option and avoid changing the decision in the future—hence resisting temptation and avoiding future self-sabotage. Common examples of this include not keeping junk food at home and not taking credit cards to a casino. Sophisticated individuals who struggle with temptation will be willing to pay to pre-commit because doing so helps them maximize their long-term well-being.

In contrast, sophisticated individuals who discount the future based on uncertainty will prefer flexibility. Their well-being improves when they can adjust their decisions in response to new information.[78] People who are aware that they face uncertain risks will be willing to pay to keep their options open so they can adapt to the new context once things become clearer. Of course, individuals who are not sophisticated (i.e., who do not recognize the underlying dynamic) will be unlikely to pay either for precommitment or for flexibility—they will always prefer the larger payment.

If people are given a choice between an option for pre-commitment and for flexibility, their decisions will give insight into which bias motivates their behavior.

## III. EXPERIMENT: TESTING TEMPTATION VS. UNCERTAINTY

This Part presents an online field experiment that tested whether the choice reversals observed in privacy behavior are explained by present bias (temptation) or by responses to uncertainty. Participants were asked to choose between disclosing personal information in exchange for immediate rewards or waiting for delayed rewards under three different conditions. By isolating the discounting mechanism at play, the design of this study allows for comparison between time-based and risk-based explanations of privacy decisions. The results support uncertainty-based discounting as a mechanism explaining privacy decisions under informational asymmetry.

---

[76] Ted O'Donoghue & Matthew Rabin, *Doing It Now or Later*, 89 AM. ECON. REV. 103, 103–04 (1999) (presenting a model of time-inconsistent preferences which distinguishes sophisticated individuals, who anticipate their future self-control problems, from naïve individuals).

[77] *Id.*; *see also* Rogers et al., *supra* note 75, at 2065.

[78] Sophisticated individuals are those who are aware of the reason for the delay-dependent discounting, while naïve individuals are those who are not.

### A. EXPERIMENT SETTING AND SAMPLE

The design builds on experimental literature in economics that distinguishes between two types of delay-dependent discounting: dynamically inconsistent (driven by temptation or present bias) and dynamically consistent (driven by uncertainty).[79] An online field experiment was designed to distinguish these mechanisms in the context of privacy decisions.[80]

Participants in the experiment were presented with a series of choices aimed at measuring their preference for pre-commitment or flexibility. A preference for pre-commitment shows that a person's behavior is driven by present bias—meaning they anticipate that their future self might make a different decision under temptation.[81] A preference for flexibility, by contrast, shows that the person is responding to uncertainty about future outcomes.[82] Instead of facing one choice between two alternative payments—as in privacy paradox experiments—participants in this study made two decisions at different times: one during a Qualtrics survey and another later over email.

Since participants did not need to interact with one another, the study could be performed online. This helps reduce external validity concerns often raised about lab-based experiments for online behavior. Participants were recruited to complete a short survey hosted and distributed by Qualtrics.[83] The survey collected basic

---

[79] Casari, *supra* note 75, at 118, 127.

[80] The difference in how people behave between the "now vs. later" scenario and the "later vs. even later" scenario is central to identifying hyperbolic discounting. People are much more likely to reverse their preferences when one of the options is immediate, suggesting that temporal proximity distorts risk perception and preference stability.

[81] Ariely & Wertenbroch, *supra* note 62, at 222–23.

[82] Marco Casari & Davide Dragone, *Choice Reversal Without Temptation: A Dynamic Experiment on Time Preferences*, 50 J. RISK & UNCERTAINTY 119, 135–36 (2015); Fernandez-Villaverde & Mukherji, *supra* note 73, at 10. *See generally* Marco Casari & Davide Dragone, *On Negative Time Preferences*, 111 ECON. LETTERS 37 (2011).

[83] A representative sample of email addresses was collected by Qualtrics, and respondents were contacted over email after the initial interaction on the Qualtrics platform. Regarding the sample size, Qualtrics' standard suggestion when the population surveyed is the general American population is that sample size be determined by: Sample Size = $(\text{Z-score})^2 * \text{StdDev}*(1-\text{StdDev}) / (\text{margin of error})^2$. A standard 95% confidence level, .5 standard deviation, and a margin of error (confidence interval) of +/- 5%, would give an ideal sample size of 385 respondents: Sample Size = $((1.96)^2 \times .5(.5)) / (.05)^2 = 385$. Other similar experiments had equivalent

demographic information (age, gender, race, postal code, and level of education), as well as participants' email address and favorite beverage at Starbucks, the store for which they received a voucher later.[84] After completing the survey portion of the study, participants were offered a choice between different types of vouchers as compensation. This choice varied across treatments.

The platform obtained 357 valid responses, distributed as follows: 119 in the control group, 118 in treatment 1, and 120 in treatment 2. Participants were located throughout the United States and spanned a range of ages. Qualtrics automatically excluded responses completed in less than 50 seconds. During data cleaning, I manually removed 8 records due to false email addresses (to which the email at the end of the experiment bounced). Participants had an incentive to provide correct email addresses because the vouchers were distributed via email. Re-including the excluded entries did not alter the results.

### B. Design and Treatments: Pricing Precommitment and Flexibility

Each participant in the study was asked to make a series of choices between two types of Starbucks vouchers. One payment option offered a higher-value voucher accompanied by a privacy loss (the no-privacy voucher, $V_n$). The other payment option offered a lower-value voucher with the avoidance of privacy loss (the privacy voucher, $V_p$). The higher-value ($V_n$) option may appear more attractive financially, but its overall value depends on how much each person values their privacy: the no-privacy-loss payoff might be larger because of the value attributed to protecting one's privacy.[85]

The $V_n$ voucher was a $7 gift card from Starbucks, but it came with the condition of accepting that the person's name and the Starbucks coffee they listed as their favorite would be published on a promotional website. The $V_p$ voucher, by contrast, was a $5 Starbucks gift card without the disclosure. The disclosure was designed to be nonsensitive: none of the collected demographic

---

(or slightly smaller) sample sizes. *See, e.g.*, Acquisti et al., *supra* note 6, at 260–66 (obtaining significant results with 349 respondents).

[84] All personal information disclosed is non-sensitive (that is, respondents will not face real-world consequences such as discrimination or social reprehension from it) and is related to a commercial interaction to take place.

[85] *See* Tesary Lin, *Valuing Intrinsic and Instrumental Preferences for Privacy*, 41 MKTG. SCI. 663, 668–71 (2022) (presenting a formal framework that considers intrinsic value of privacy); Ignacio Cofone, *Nothing to Hide, but Something to Lose*, 70 U. TORONTO L.J. 64, 70–80 (2020) (presenting a model that considers intrinsic value of privacy that can be compared to external rewards).

information was shared and participants were informed of this so that they would not expect any real-world consequences from information that would be made public—doing so better captured the intrinsic value they place on privacy.

In the control group (baseline treatment), by the end of the survey portion (at $t_1$) participants were given a choice: they could either (a) pick any of the two vouchers ($V_n$ and $V_p$) right away or (b) postpone the decision and choose between them after they receive a follow-up email a week later (at $t_2$), which would contain further information about the context of the disclosure. That follow-up email provided them with more information about the promotional website that would publish the information and the format in which it would be published.[86] Importantly, the timing of the reward remained the same for everyone. So, because the payment was delayed to the reception of the email, the choice during the survey portion ($t_1$) took place in a low-temptation environment. Telling participants that they would receive this new information was important for isolating the discounting mechanism because the delay alone should not make a difference for participants under uncertainty-based discounting unless they expect more information to arrive.

Regardless of their decision at the survey portion, participants received the email with the website details (at $t_2$). That email either prompted them to make a choice between vouchers with a one-week period (if they had not made the choice yet) or reminded them of the choice they made (if they had). One week after that first email (at $t_3$) all vouchers were sent to participants via email. The design of the study is illustrated in Figure 1. The first row represents $t_1$, the second and third row represent $t_2$, and the last row represents $t_3$.

[86] *See* Appendix A *infra* ("The website will just list the names of participants of this survey who chose to be part of it, and what is their favorite beverage. It will not be linked directly to any company website.")

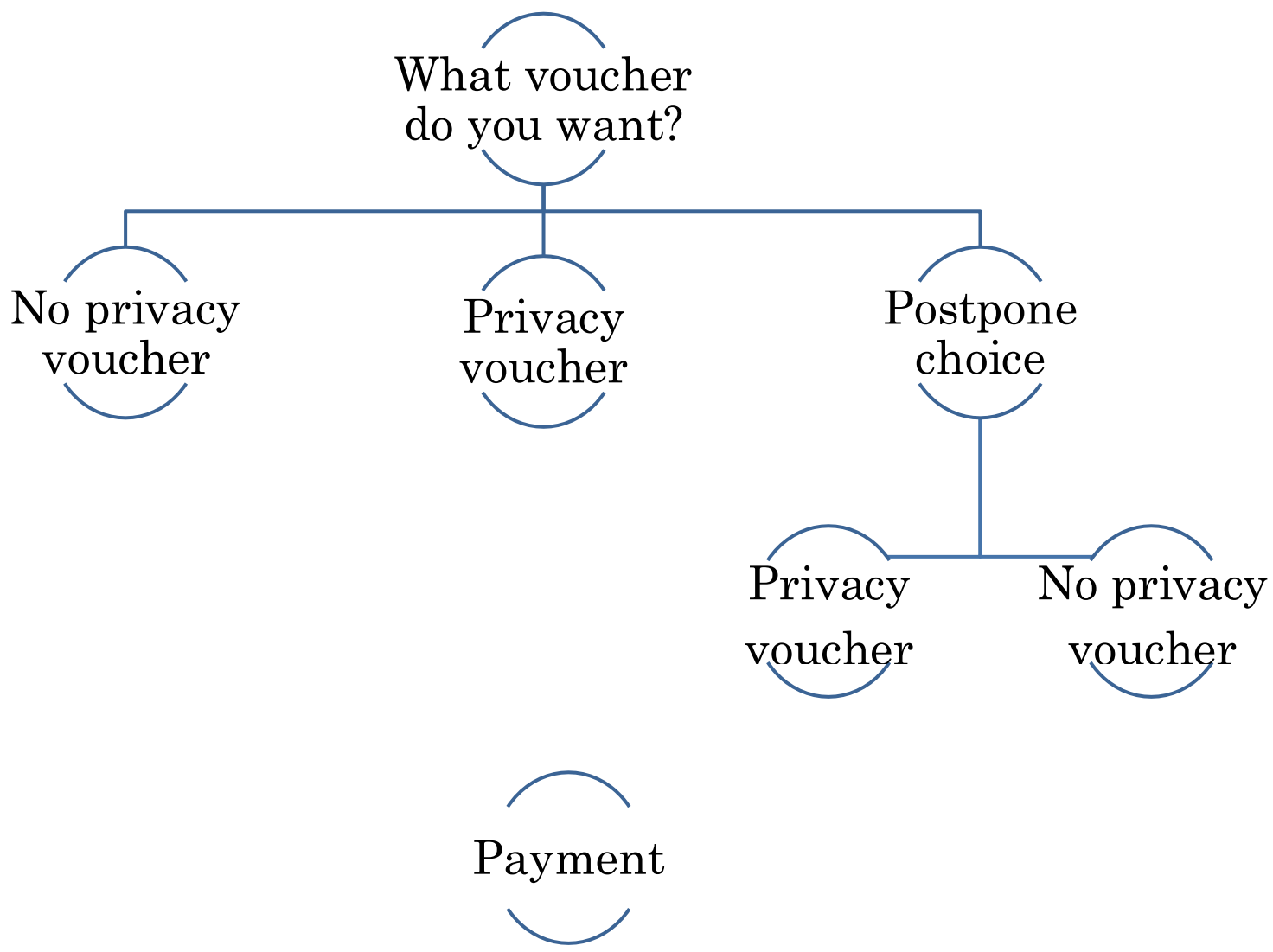

Figure 1. Illustrates the design common across treatments (dollar amounts vary between treatments).

Two variants of the voucher choice, which was available to all participants at $t_1$ and $t_2$, were introduced under separate treatments.

Treatment 1 made choosing between vouchers later (i.e., flexibility) slightly costly. This was done by lowering the value of the voucher options available later ($t_2$) by \$1, so that participants had to pay to keep their options open at $t_1$ (so that $V'_p<V''_p$ and $V'_n<V''_n$). Participants in this treatment had three options: (a) pre-commit at $t_1$ to a \$5 privacy-preserving voucher ($V_p$), (b) choose at $t_1$ a \$7 no-privacy voucher ($V_n$), and (c) choose a week later at $t_2$ between a \$4 privacy-preserving voucher ($V''_p$) and a \$6 no-privacy voucher ($V''_n$).

Flexibility being only available at a cost in Treatment 1 allows one to measure whether participants are willing to pay for it. Maintaining flexibility should be helpful for participants discounting based on uncertainty since they can decide after receiving more information. Participants who responded primarily to uncertainty should find this option valuable.

Treatment 2 made choosing between vouchers earlier (i.e., pre-commitment) costly by \$1. This was done by slightly lowering the value of the voucher options available at the initial decision point (so that $V'_p<V_p$ and $V'_n<V_n$). In this treatment, participants chose between three options: a \$4 voucher that protects their privacy at $t_1$ ($V'_p$), a \$6 voucher that involved disclosure at $t_1$ ($V'_n$), and delaying the decision until a week later ($t_2$) choosing over email between a \$5 privacy-preserving voucher ($V_p$) and a \$7 no-privacy voucher ($V_n$).

Pre-commitment having a small cost in Treatment 2 allows one to measure whether participants are willing to pay for it. If some participants expect that they might be tempted to take the higher-value voucher when the payments are close and regret giving up their privacy (i.e., they know that they might give in to temptation when the payment is immediate), this mechanism would help them: they could choose to lock in the privacy-protecting voucher ($V_p$) earlier, instead of waiting and risking that temptation might lead them to choose the other one ($V_n$) when the reward is in front of them.[87] Participants facing temptation would see value in locking in a privacy-protecting option before the temptation arises and should be willing to pay for pre-committing to privacy.[88]

### C. RESULTS: FLEXIBILITY BEATS PRECOMMITMENT IN INDIVIDUAL CHOICES

The experiment is designed to compare behavior across treatments to measure whether participants value pre-commitment and flexibility. The difference between the proportion of participants who chose to delay the decision in treatment 1 (where flexibility is costly) and the proportion who do so in the control group shows how much participants valued flexibility. Conversely, the difference between the proportion of participants who chose $V_p$ in treatment 2 (where pre-commitment is costly) and the proportion who chose $V_p$ in the control group (where pre-commitment is not costly) shows how much participants valued pre-commitment.

If participants value pre-commitment more than flexibility (WPC>WPF), this would suggest that their choice is driven primarily by temptation—they want to lock in their privacy choice before facing temptation when the payoff is in front of them. If the opposite is true (WPC<WPF), this would suggest that, as participants want to wait for more information before deciding, their decisions are driven primarily by uncertainty.

The results are presented in Table 3 and Figure 2, below.

87 O'Donoghue & Rabin, *supra* note 76, at 105–07 (acknowledging that, among individuals that face temptation, sophisticated individuals will prefer commitment devices that restrict their future choices).

88 *Id.* at 111. Subjects may choose the lower-value privacy voucher at time T as a form of pre-commitment, anticipating that they will be more tempted by the higher-value but privacy-invasive option at T+1. This aligns with models of sophisticated self-control, in which individuals are aware of their tendency to make short-sighted choices in the future and take steps to constrain their future options.

| Group | Now, $V_p$ (number) | Now, $V_n$ (number) | Later (number) | Now, $V_p$ (percent) | Now, $V_n$ (percent) | Later (percent) |
|---|---|---|---|---|---|---|
| **Control (n=119)** | 30 | 63 | 26 | 25.21008 | 52.94118 | 21.84874 |
| **Treatment 1 (n=118)** | 24 | 77 | 17 | 20.33898 | 65.25424 | 14.40678 |
| **Treatment 2 (n=120)** | 12 | 35 | 73 | 10 | 29.16667 | 60.83333 |

Table 3: Choices by treatment

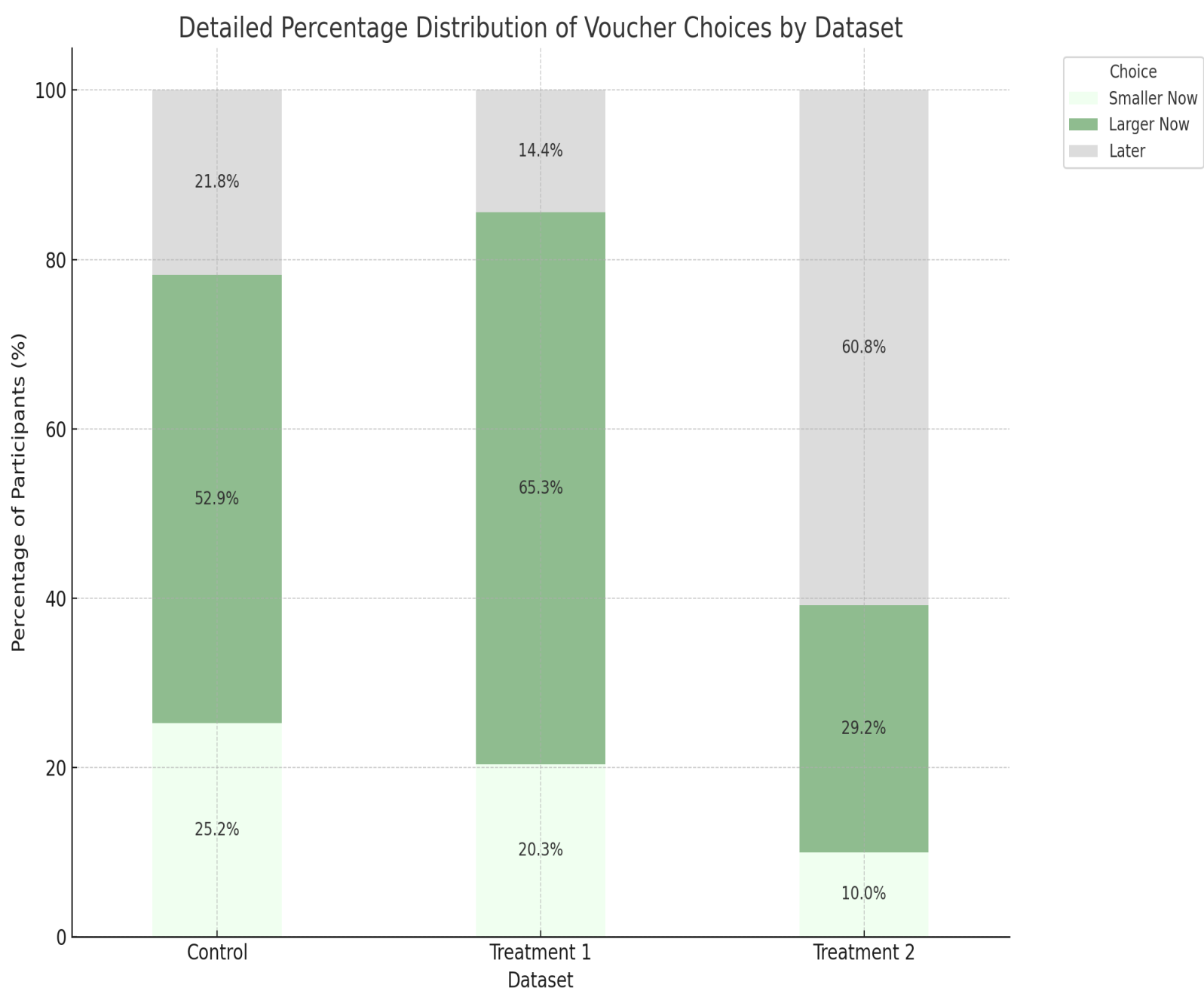

Figure 2. Breaks down percentage results by voucher

Note that most people in the control group chose one of the two "now" options.[89] That might be because flexibility, even in the control, is not entirely costless: some participants may choose immediately in the baseline condition to resolve the issue and "get it over with" if they feel that the effort of re-engaging later is undesirable (even if technically free).

To answer whether participants on average preferred pre-commitment or flexibility, it is helpful to do two things. The first is to start by comparing the relative sizes of participants willing to pay $1 for their preferred choice format.[90] More data volume might establish a clearer relationship as the numbers for this comparison were small. Pay-for-pre-commitment, captured by the number of low voucher "now" choices in treatment 2, is 10%; pay-for-flexibility, captured by the number of "later" choices in treatment 1, is 14.4%. That means the ratio, in those small numbers, was approximately 1:1.44: at the $1 price point, participants were about 44% more likely to pay $1 for flexibility than they were to pay $1 for precommitment.

More informative is the second comparison: evaluating participants' shifts in choices, indicating WPC and WPF.[91] Penalizing participants for making their choices flexible led to a modest and statistically insignificant deviation from the baseline (34% decrease). On the other hand, penalizing participants for pre-committing generated a large and statistically significant deviation from the baseline (60% decrease).[92] This means that while only a minority paid for either pre-commitment or flexibility outright (10% and 14.4%), many (60.8%) chose "later" when "now" was disadvantaged.[93] That indicates a flexibility-preferring pattern.

Risk ratios showed that participants responded more to making pre-commitment expensive (they flock to "later") than to making flexibility expensive. When the treatment made choosing "now" worse (Treatment 2), far more people chose to wait: the share picking "later" jumped by 38.98 percentage points from 21.85% in the baseline to 60.83%. In statistical terms, that is a risk ratio of 2.78 (95% CI 1.93–4.03).[94] That shift is large: the odds of choosing "later" were about three times higher than at baseline. By contrast, when the treatment made waiting worse (Treatment 1), the "later" share decreased only by 7.44 percentage points from 22% to 14%, which is small: the risk ratio against the baseline is 0.65 (95% CI 0.38–1.15).[95] People, in other words, moved toward flexibility when deciding earlier was slightly costlier, but they did not move away from flexibility nearly as much when waiting was costlier. The first ratio is over four times the size of the second one, which reveals a preference for keeping options open.

---

[89] For context, in Acquisti, John & Loewenstein's study using $10 and $12 Visa gift cards, in the treatment where respondents were given a neutral choice and the $10 card was listed before the $12 card, 57.8% of respondents chose the larger card, and when the larger card was listed first, 73.3% respondents chose it. *See* Acquisti et al., *supra* note 6, at 260–66.

[90] Done in a two-proportion z-test on the WPC and WPF numbers (two-tailed).

[91] Comparing each of the two sample proportions using a 2-sample z-test. *See* Appendix B *infra*.

[92] *See* Appendix B *infra*.

[93] About a third of those who chose "later" when "now" was disadvantaged (20.3%) pre-committed to their privacy choice even when disadvantaged.

The difference between the proportion of participants who chose $V_p$ in treatment 2 (where pre-commitment is costly) and the proportion who chose $V_p$ in the control group, where pre-commitment is not costly, was also statistically significant at a 60.2% decrease. In other words, people significantly moved away from pre-commitment when it was costly. When flexibility was costly at treatment 1, there was a statistically insignificant difference with fewer people choosing to pre-commit than in the control.[96]

Next, one might interpret thresholds on net willingness to pay for flexibility using three price points ($p$=-1; 0; 1). From the results, one would infer 14.4% of participants had WPF ≥ $1 (they chose "later" even when it costed $1 in Treatment 1), while 7.4% had a WPF [0, $1) (they chose "later" when free, but not when it costed $1; Control vs Treatment 1), 39% had a net WPF [−$1, 0) (switch to "later" when "now" was penalized by $1; Control vs Treatment 2), and only 10% had WPF < −$1 (stick with "now" even when penalized by $1; Treatment 2). This distribution is consistent with a meaningful but heterogeneous demand for flexibility: many are near the margin and tilt toward delaying their decision when incentives nudge them.

If one were to account for the nuisance cost of delaying the decision and spending more time on deciding over the voucher, the numbers would change slightly. In the baseline condition, some participants may not be motivated to delay the decision because they do not yet know how much the additional information will change their valuation. That is, unless they expect the website details to alter their perception of the privacy risk, they may not see the benefit of delaying. Without a reason to expect that the value of waiting will be high, people may default to a decision in the moment. Substantively, the conclusion is unchanged: a minority would pay for flexibility, and many shift to waiting when "now" is slightly disadvantaged. While considering the nuisance cost of waiting would increase the range of people with WPF, it is difficult to estimate by how much.[97]

---

[94] Odds ratio = 5.56 (95% CI 3.15–9.81).

[95] Odds ratio = 0.60 (95% CI 0.31–1.18).

[96] 0.048711 proportion difference (19.32% decrease); z-value= 0.9; p= 0.3714.

These patterns indicate that participants, on average, valued flexibility more than pre-commitment in disclosure choices. The results suggest that, to the extent that participants discount future privacy, they do so primarily (although not necessarily exclusively) due to structural uncertainty.

The results did not vary by gender, age, ethnicity, level of education, region, or location (rural versus urban).[98] A relevant distinction is that the experiment tested for temptation vs uncertainty *for data disclosure*. In other words, it compares whether people face structural uncertainty with whether they feel tempted *to disclose their personal data*. This does not rule out any other behavioral biases or, even, people being tempted *to use an application* which then collects data from them in a context of structural uncertainty. Temptation to use applications and addiction, for which there is abundant empirical evidence,[99] are compatible with the structural uncertainty account.[100]

## IV. INTERPRETATION AND POLICY IMPLICATIONS

The findings align with concerns raised by regulators and advocates: that privacy harms arise not from poor choices, but from opaque data practices beyond people's control. In a world where individual well-being depends on how companies use personal data, recognizing uncertainty as the driver of privacy behavior shifts the focus from blaming people to holding systems accountable.

---

[97] To account for the nuisance cost of waiting, one could interpret the cutoff estimates as net WPF. With a constant nuisance cost of delaying ($\delta$), gross WPF = net WPF + $\delta$. The thresholds would shift by $\delta$. For example, with $\delta$ = \$1 these become \$2, \$1, and \$0.

[98] In Treatment 2, more women (43.08%) chose "now" compared to men (34.55%), and more men (65.45%) deferred the choice to "later" compared to women (56.92%). There's a seeming difference between people with postgraduate education in Treatment 2 but that might be driven by the small sample size of that group in that treatment (20).

[99] *See generally* Maèva Flayelle, Damien Brevers, Daniel L. King, Pierre Maurage, José C. Perales & Joël Billieux, *A Taxonomy of Technology Design Features that Promote Potentially Addictive Online Behaviours*, 2 NATURE REVS. PSYCH. 136 (2023).

[100] *See, e.g.*, Matthias Sutter, Martin G. Kocher, Daniela Glätzle-Rützler & Stefan T. Trautmann, *Impatience and Uncertainty: Experimental Decisions Predict Adolescents' Field Behavior*, 103 AM. ECON. REV. 510, 525–28 (2013).

### A. CALLS FOR NUDGES AND TRANSPARENCY

If privacy decisions were driven by people being tempted to disclose more information than they would like to, people's optimal strategy would be pre-commitment. In other contexts, such as dieting, addiction, or saving, people benefit from mechanisms that restrict future choices to avoid succumbing to short-term temptation. Such an interpretation of the privacy paradox would favor paternalistic or libertarian-paternalistic interventions, or nudges designed to align behavior with stated privacy preferences.[101] Regulation that incorporates such interpretation would provide tools for people to pre-commit not to disclose personal information (in addition to any pre-commitment from engaging with the services themselves, for example to counteract addictive design, which is isolated in this experiment as everyone was engaging with the setup).[102]

The structural uncertainty account that interprets that people discount based on unknown risk reverses this idea and leads to the conclusion that policy should provide people with increased transparency and flexibility for their privacy choices. Table 2, below, summarizes this contrast.

| | **Bias-based Discounting** | **Uncertainty-based Discounting** |
|---|---|---|
| **Intervention goal** | Modify preference or pre-commit | Clarify risk, allow reversibility |
| **Legal implication** | Cooling-off periods, nudges, and pre- commitment tools | Transparency rights, RTBF |
| **Regulatory paradigm** | Behavioral paternalism | Data protection law |

Table 4: Competing implications

To evaluate the intuitiveness of the structural uncertainty account supported by the results outlined above,[103] one can examine what people and consumer associations have demanded for privacy: pre-commitment mechanisms or flexibility.

Privacy advocacy groups and consumer associations overwhelmingly focus on the lack of transparency in personal data processing—highlighting issues like hidden profiling, unknown data brokerage, and obscure third-party transfers.[104] The European Consumer Organization, for instance, calls targeted advertising "a hidden side of the data economy,"[105] and has long said that people "are sleep-walking in a world without privacy. They do not realize their data is being collected and processed."[106] The Electronic Frontier Foundation refers to a "disturbing lack of transparency" about how data is collected, shared, and used.[107] Privacy International repeatedly describes people's privacy decision-making context as a "hidden data ecosystem."[108]

[101] Alessandro Acquisti, *Nudging Privacy: The Behavioral Economics of Personal Information*, 7 IEEE SEC. & PRIV. 72, 74 (2009).

[102] Alternatively, a regulation aiming to do this could create a system of reward substitution—paying to avoid disclosure, charging to disclose, creating guilt, imposing additional obstacles, etc.

[103] *See supra* Parts II.C, III.C.

Similarly, a popular objection to targeted advertising is visceral, with many describing it as "creepy."[109] This term reflects emotional discomfort arising from the mismatch between what people expect and what is revealed by data uses.[110] It fits the characterization that people are uncertain of which companies have information about their interests until they are shown targeted advertisements.[111] In this context, consumer reactions to privacy harms rarely resemble regret over failed self-restraint not to disclose in their engagement with those apps. Instead, they reflect surprise or frustration at unexpected uses of data that people did not foresee at the time of data collection.[112] At least from casual empiricism, similarly, social network users do not typically promise themselves to stop sharing their personal information online and fail in their efforts,[113] as do people who face temptation in other contexts, such as when they are dieting or quitting smoking.[114] By contrast, surveys indicate that people do not understand how their data is collected and used.[115] And reporting shows that people are often shocked to find how advertisers use their information to show them ads on topics they recently discussed with others or inquired about.[116]

---

[104] *See, e.g.*, FED. TRADE COMM'N, A LOOK BEHIND THE SCREENS: EXAMINING THE DATA PRACTICES OF SOCIAL MEDIA AND VIDEO STREAMING SERVICES (2024), https://www.ftc.gov/reports/look-behind-screens-examining-data-practices-social-media-video-streaming-services [https://perma.cc/82LE-RKCY]; U.S. GOV'T ACCOUNTABILITY OFF., GAO-22-106096, CONSUMER DATA: INCREASING USE POSES RISKS TO PRIVACY (2022), https://www.gao.gov/products/gao-22-106096 [https://perma.cc/BDW5-3T49]; FED. TRADE COMM'N, DATA BROKERS: A CALL FOR TRANSPARENCY AND ACCOUNTABILITY (2014), https://www.ftc.gov/system/files/documents/reports/data-brokers-call-transparency-accountability-report-federal-trade-commission-may-2014/140527databrokerreport.pdf [https://perma.cc/V3KX-5NWJ].

[105] Email from Ursula Pachl, BEUC Deputy Dir. Gen., Civil Society Organisations Take Action Against Widespread Commercial Surveillance by Adtech Industry, to Elizabeth Denham, Chair of the Glob. Priv. Assembly (Apr. 21, 2020), https://www.beuc.eu/sites/default/files/publications/beuc-x-2020-027_action_against_widespread_commercial_surveillance_by_adtech_industry.pdf [https://perma.cc/J7B4-SQQQ].

[106] Matt Warman, *EU Fights "Fierce Lobbying" to Devise Data Privacy Law*, THE TELEGRAPH (Feb. 9, 2012, 7:00 AM), https://www.telegraph.co.uk/technology/internet/9069933/EU-fights-fierce-lobbying-to-devise-data-privacy-law.html [https://archive.ph/kNKuk].

[107] Lena Cohen, *FTC Report Confirms: Commercial Surveillance is Out of Control*, ELEC. FRONTIER FOUND. (Sept. 26, 2024), https://www.eff.org/deeplinks/2024/09/ftc-report-confirms-commercial-surveillance-out-control [https://perma.cc/RR3M-YRJ6].

[108] *Challenge to Hidden Data Ecosystem*, PRIV. INT'L, https://privacyinternational.org/legal-action/challenge-hidden-data-ecosystem [https://perma.cc/59PM-EAB5].

[109] *See, e.g.*, Audrey Schomer, *Most Consumers Are Creeped Out by Ads That Follow Them Across Devices*, EMARKETER (July 23, 2021), https://www.emarketer.com/content/most-consumers-creeped-out-by-ads-

---

that-follow-them-across-devices [https://perma.cc/7EJC-FXYC] (reporting about two-thirds of respondents said ads that "follow them" across devices are creepy).

[110] Paul M. Schwartz & Daniel J. Solove, *The PII Problem: Privacy and a New Concept of Personally Identifiable Information,* 86 N.Y.U. L. REV. 1814, 1853 (2011); Blase Ur, Pedro Giovanni Leon, Lorrie Faith Cranor, Richard Shay & Yang Wang, *Smart, Useful, Scary, Creepy: Perceptions of Online Behavioral Advertising*, *in* SYMPOSIUM ON USABLE PRIVACY AND SECURITY 7, 11 (2012).

[111] Schwartz & Solove, *supra* note 110, at 1853 ("As for transparency, behavioral marketing takes place today in a multi-channel process about which individuals generally receive scant information about the data that organizations collect about them or how that information is used to shape interactions with them."); *see also* Tsai et al., *supra* note 31, at 260–61.

[112] DANIEL J. SOLOVE, THE DIGITAL PERSON: TECHNOLOGY AND PRIVACY IN THE INFORMATION AGE 50–51 (Jack M. Balkin & Beth Simone Noveck eds., 2004).

[113] *Id.* at 44–49.

[114] Cass R. Sunstein & Richard H. Thaler, *Libertarian Paternalism is Not an Oxymoron*, 70 U. CHI. L. REV. 1159, 1162–63 (2003) (describing patterns of temptation and self-control failure in traditional behavioral contexts like dieting and saving).

[115] *See* McClain et al., *supra* note 3 (finding 67% of U.S. adults say they understand little to nothing about what companies do with their personal data).

[116] Claire M. Segijn, Joanna Strycharz, Anna Turner & Suzanna J. Opree, *"My Phone Must be Listening!": Peoples' Surveillance Beliefs Around Devices "Listening" to Offline Conversations in the US, the Netherlands, and Poland*, 12 BIG DATA & SOC'Y, Apr.–June 2025 (finding that among U.S. participants who reported seeing conversation-related ads,

The experimental evidence from prior research outlined above also supports the need for transparency and flexibility measures.[117] For example, experimental evidence shows that reactions to privacy choices change depending on the choices' complexity.[118] Furthermore, it shows that, when information about privacy is visible, people choose higher privacy protections.[119] This behavior suggests that what appears to be irrational or inconsistent behavior might instead be a response to people facing an uncertain decision-making scenario with hidden risks, since irrationality would prevent people from incorporating the new information into their decisions.[120]

At a more general level, the display of different valuations for different types of information and the reaction to changes in context and accessibility point to this account. In contrast to people making uncertainty-driven choices, people facing temptation already have the relevant information about the decision context, so they discount independently of new information they receive.[121] But experimental evidence shows that privacy decisions change when new information becomes salient.[122]

### B. UNCERTAIN DATA HARMS

Privacy choices differ from choices typically captured in behavioral experiments that focus on temptation (present-bias) in two ways. The first is that those experiments usually involve choosing between two positive outcomes: receiving money now or receiving more money later.[123] In contrast, privacy decisions involve

---

electronic eavesdropping was the top explanation at 47.2%); Bree Fowler, *Is Your Smartphone Secretly Listening to You?*, CONSUMER REPS. (July 10, 2019), https://www.consumerreports.org/smartphones/is-your-smartphone-secretly-listening-to-you [https://perma.cc/X8LW-UWUX] (reporting 43% of American smartphone owners said they believe their phone is recording their conversations without permission).

[117] *See supra* Part I.B.

[118] John et al., *supra* note 34, at 868.

[119] Gideon et al., *supra* note 30, at 139–41; Tsai et al., *supra* note 31, at 263.

[120] *See* Acquisti & Grossklags, *supra* note 15, at 27 (arguing that many deviations in privacy behavior are neither unreasonable nor truly irrational; rather, they reflect sensible heuristics given uncertainty); *see also* Epper et al., *supra* note 50, at 185–92*;* Story et al., *supra* note 50, at 268-71.

[121] *See generally* RUSSELL SAGE FOUND., CHOICE OVER TIME (George Loewenstein & Jon Elster eds., 1992).

[122] *See supra* Part I.B.

[123] Internet users face losses instead of gains as in most of the hyperbolic discounting literature. *See generally* O'Donoghue & Rabin,

avoiding a negative outcome, such as assessing the negative repercussions of missing out on using a digital product or service against long-term potential privacy harms.[124]

The second and most relevant difference is the uncertainty of the outcomes. In most behavioral experiments (because participants choose between monetary payoffs at different times), participants know exactly how much money they will receive with each option and are either certain that it will be delivered or know the exact probability that it will. Similarly, in real-life situations in which people face temptation (i.e., they hyperbolically discount based on behavioral biases), such as choosing an unhealthy but tasty snack option over a healthy, less tasty snack option, people generally understand the risks involved (i.e., they have a notion of the payoffs' sizes and their probability). If they choose the unhealthy but tastier option, it is reasonable to interpret the decision as influenced by temptation because they knew about the risk beforehand. If they had not known the differential health effects of the choices, the decision would not have been based on how they discount the future.

Privacy decisions are not like that.[125] A privacy harm, which materializes the risk in these decisions, can occur with an unknown probability at each moment.[126] When people disclose personal information online, they cannot know the probability of the delayed penalty (privacy harm), as they are unaware of the risk.[127] Every time companies share or sell a user's personal information, or every time a company is hacked, the risk of privacy harm to that person

---

*supra* note 76 (illustrating this difference). While gains are preferred now better than later, losses are preferred later better than now. This is notwithstanding the fact that people seem to discount losses with a lower discount rate than the one they use to discount gains. *See* Thaler, *supra* note 44, at 205.

[124] Acquisti et al., *supra* note 8, at 444, 451.

[125] *See* SIVA VAIDHYANATHAN, THE GOOGLIZATION OF EVERYTHING (AND WHY WE SHOULD WORRY) 84 (2011).

[126] *See* Solove & Citron, *supra* note 36, at 741; Citron & Solove, *supra* note 63, at 816–17.

[127] Strandburg, *supra* note 71, at 130–32; Ignacio Cofone & Adriana Robertson, *Consumer Privacy in a Behavioral World*, 69 HASTINGS L.J. 1471, 1489–90 (2018); Joseph Turow, Chris Jay Hoofnagle, Deirdre K. Mulligan, Nathaniel Good & Jens Grossklags, *The Federal Trade Commission and Consumer Privacy in the Coming Decade*, 3 I/S: J.L. & POL'Y FOR INFO. SOC'Y 723, 733 (2007), https://repository.upenn.edu/handle/20.500.14332/2348; Solove & Citron, *supra* note 36, at 757; Daniel J. Solove, *Privacy Self-Management and the Consent Dilemma*, 126 HARV. L. REV. 1880, 1883–88 (2012).

increases.[128] This includes both legitimate and illegitimate transfers of information. Keeping other conditions stable, the more places someone's personal information ends up in, the more likely it is to be exposed or misused, but people do not have control after the initial moment of collection. The externalities in data trading[129] mean that, while data collectors and intermediaries or intermediaries and advertising companies agree on these exchanges, people face risks of harm from each one.

The same is true for data processing. When a company uses someone's personal information, the information is out of that person's range of control but data practices can still impact them negatively.[130] Since companies do not face all costs, they have an incentive to overuse and over-trade user information.[131] This is aggravated by the fact that people often do not know about harmful practices until it is too late, so they have no opportunity to "discipline" companies that take on risky uses by engaging with them less.[132]

In privacy, choices' potentially large negative payoffs do not occur with certainty.[133] Therefore, a discount rate inferred only from observed behavior will conflate both discounting mechanisms: the discount for the penalty's delay and the discount for its probability

---

[128] *See* Acquisti et al., *supra* note 8, at 449 (explaining how data proliferation increases vulnerability and users cannot retrieve or limit it once shared.).

[129] *See* Hal R. Varian, *Economic Aspects of Personal Privacy*, *in* CYBER POLICY AND ECONOMICS IN AN INTERNET AGE 127 (William H. Lehr & Lorenzo M. Pupillo eds., 2002). *See generally* Kenneth C. Laudon, *Markets and Privacy*, 39 COMMC'NS ACM 92 (1996).

[130] IGNACIO COFONE, THE PRIVACY FALLACY: HARM AND POWER IN THE INFORMATION ECONOMY 59–62 (2023).

[131] Jay Pil Choi, Doh-Shin Jeon & Byung-Cheol Kim, *Privacy and Personal Data Collection with Information Externalities*, 173 J. PUB. ECON. 113, 117, 120 (2019) (showing that negative privacy externalities lead to socially excessive collection and usage of data); John Hagel III & Jeffrey F. Rayport, *The Coming Battle for Consumer Information*, 75 HARV. BUS. REV. 53, 53–65 (1997).

[132] PETER P. SWIRE & ROBERT E. LITAN, NONE OF YOUR BUSINESS: WORLD DATA FLOWS, ELECTRONIC COMMERCE, AND THE EUROPEAN PRIVACY DIRECTIVE 8 (James Schneider ed., 1998).

[133] *See* Solove & Citron, *supra* note 36, at 741; Citron & Solove, *supra* note 63, at 816–17.

of occurrence.[134] Any perceived risk would alter the discounting for delay.[135]

Hence, even if someone had full information about costs and benefits at the moment of making a privacy choice, they would have to base the decision on an uncertain risk. The risk of privacy harm is not dependent on user behavior alone, but also depends on the subsequent behavior of companies that acquire their personal data.[136] This leads to an impossibility in making an optimal decision and, as the last part showed, leads to a discount function that will produce a choice reversal.

Moreover, we know from behavioral science that uncertainty over risk is not only determined by the unknown externalities. In addition to this objective uncertainty, people have subjective uncertainty due to high information costs.[137] People, for example, often do not understand privacy policies;[138] nor do they understand how to use privacy protection tools.[139] Managing privacy risks and protecting one's privacy online requires knowledge and technical skills that few have,[140] so many do not know how to take measures to protect their privacy. More than half of Americans believe that the mere existence of a privacy policy means that companies cannot trade their personal data.[141] Many mistakenly believe that the average probabilities of specific data harms, such as identity fraud and identity theft, are lower than they actually are.[142]

---

[134] Epper et al., *supra* note 50, at 186; *see also* Blackburn & El-Deredy, *supra* note 64.

[135] Sozou, *supra* note 52, at 2018; Halevy, *Time Consistency*, *supra* note 61, at 1148.

[136] Cofone, *supra* note 130, at 59–62.

[137] *See* Acquisti & Grossklags, *supra* note 15, at 26–27.

[138] Joel R. Reidenberg, Travis Breaux, Lorrie Faith Cranor, Brian French, Amanda Grannis, James T. Graves, Fei Liu, Aleecia McDonald, Thomas B. Norton, Rohan Ramanath, N. Cameron Russell, Norman Sadeh & Florian Schaub, *Disagreeable Privacy Policies: Mismatches Between Meaning and Users' Understanding*, 30 BERKELEY TECH. L.J. 39, 40–41, 83–87 (2015); Jenny Tang, Hannah Shoemaker, Ada Lerner & Eleanor Birrell, *Defining Privacy: How Users Interpret Technical Terms in Privacy Policies*, PROC. PRIV. ENHANCING TECH. 70 (2021); Kim-Phuong L. Vu, Vanessa Chambers, Fredrick P. Garcia, Beth Creekmur, John Sulaitis, Deborah Nelson, Russell Pierce & Robert W. Proctor, *How Users Read and Comprehend Privacy Policies*, *in* HUMAN INTERFACE AND THE MANAGEMENT OF INFORMATION: INTERACTING IN INFORMATION ENVIRONMENTS 802 (Michael J. Smith & Gavriel Salvendy eds., 2007).

[139] Acquisti & Grossklags, *supra* note 46, at 9.

[140] *See* Susanne Barth, Menno D.T. de Jong & Marianne Junger, *Lost in Privacy? Online Privacy from a Cybersecurity Expert Perspective*, 68 TELEMATICS & INFORMATICS 1039, 1046 (2022) (showing that many users rely on superficial cues such as app ratings and design rather than employ technical knowledge or skills to manage privacy risks, demonstrating a lack of technical understanding); *see also* Solove, *supra* note 65, at 984 (arguing that few people have the knowledge to exercise privacy rights adequately).

Policy responses that equate discounting and dynamic inconsistency can produce harmful outcomes if people are in fact consistent.[143] Although optimal for people who are present-biased, eliminating future choices (pre-committing) is detrimental when more information is expected.[144] In the information economy, the costs to people depend on the future behavior of those processing their information, so new information is likely to arise. Data protection law, as the next part explains, should account for the fact that people may alter their data choices when experiencing a change in context.

## V. REGULATORY IMPLICATIONS: DESIGN FOR STRUCTURAL UNCERTAINTY

The structural uncertainty account suggests that privacy law should focus on providing people with tools that enable them to learn more about the context of privacy-related decisions (a response to people who discount based on an uncertain risk). This aligns with recent critiques of privacy regulation that emphasize the importance of shifting focus to systemic improvements and transparency.[145] While it is hardly possible to eliminate uncertainty through regulation, decreasing it is possible if privacy law establishes appropriate measures. Three sets of measures can do so: targeted transparency obligations; reduced reliance on privacy policies; and flexibility introduced by data control rights.

---

141 *See* Joseph Turow, *Americans and Marketplace Privacy*, *in* THE CAMBRIDGE HANDBOOK OF CONSUMER PRIVACY 151 (Evan Selinger, Jules Polonetsky & Omer Tene eds., 2018); Turow et al., *supra* note 127, at 733.

142 *See* Acquisti & Grossklags, *supra* note 15, at 30 (comparing survey data with data from the United States Federal Trade Commission, finding that over 70% of people underestimate the probabilities of identity theft).

143 *See* Azfar, *supra* note 61, at 251 ("[W]e should be careful about confusing non-constant discounting with dynamic inconsistency.").

144 Manuel Amador, Iván Werning & George-Marios Angeletos, *Commitment vs. Flexibility*, 74 ECONOMETRICA 365, 365–66 (2006).

145 *See, e.g.*, Filippo Lancieri, *Narrowing Data Protection's Enforcement Gap*, 74 ME. L. REV. 15, 17–19 (2022) (describing how structural features like market power and information asymmetries hinder data protection compliance despite regulatory frameworks like the GDPR and the California Privacy Protection Act).

### A. Enhanced Transparency Obligations

Reframing the so-called privacy paradox as a product of uncertainty shifts the regulatory focus toward the structure of the information environment. The findings of this Article show that when people appear to disclose personal data contrary to their stated preferences, it is primarily due to a rational response to poorly understood risks. In this light, reducing that uncertainty through targeted transparency obligations is a warranted intervention. The evidence provided supports calls in legal scholarship for transparency mechanisms to address information asymmetries, particularly on data flows and algorithmic inference.

Statutory frameworks in U.S. privacy law, such as the CPPA, as well as in data protection law abroad, such as the General Data Protection Regulation (GDPR), already impose a baseline set of transparency requirements on data controllers. For instance, the GDPR requires controllers to inform people about the purposes of processing, categories of data collected, retention periods, recipients of the data, and existence of automated decision-making, including profiling.[146] However, these obligations are often satisfied through dense privacy policies or general statements that fail to meaningfully reduce uncertainty about what consequences may follow from data practices.[147] Even though the GDPR requires that controllers disclose detailed information about their data practices to their users,[148] the way this information is presented often renders it ineffective at mitigating this form of uncertainty.[149] To address structural uncertainty, one should refine transparency obligations to target the lack of knowledge that is most relevant for decision-making: more than transparency over the specifics of a data practice itself, people benefit from transparency over its potential consequences.

[146] Regulation (EU) 2016/679, arts. 13–14, 2016 O.J. (L 119) 1.

[147] Peter J. van de Waerdt, *Information Asymmetries: Recognizing the Limits of the GDPR on the Data-Driven Market*, 38 COMPUT. L. & SEC. REV., Sep. 2020, at 1–2 (GDPR notices "in practice unable to mitigate" data asymmetries; consumers remain in a "vulnerable position."); Giulia Grundler, Rūta Liepina, Mariaceleste Musicco, Francesca Lagioia, Andrea Galassi, Giovanni Sartor & Paolo Torroni, *Detecting Vague Clauses in Privacy Policies: The Analysis of Data Categories Using BERT Models and LLMs*, *in* LEGAL KNOWLEDGE & INFORMATION SYSTEMS 72, 72–76, 81 (2024) (showing that GDPR privacy policies frequently include vague clauses, obstructing user understanding).

[148] Regulation (EU) 2016/679, arts. 13–14, 2016 O.J. (L 119) 1; *see also* INFO. COMM'R OFF., *Right to Be Informed*, https://ico.org.uk/for-organisations/uk-gdpr-guidance-and-resources/individual-rights/individual-rights/right-to-be-informed [https://perma.cc/N3LG-WFZS]; DATA PROT. COMM'N, *The Right to be Informed (Transparency) (Article 13 & 14 GDPR)*, https://www.dataprotection.ie/en/individuals/know-your-rights/right-be-informed-transparency-article-13-14-gdpr [https://perma.cc/3DSK-DKLU].

[149] *See* van de Waerdt, *supra* note 147.

The first step is inferential transparency: disclosures should more precisely explain the types of inferences that may be drawn from personal data.[150] People often understand that their data may be collected but remain unaware of how it may be aggregated to generate sensitive inferences (for example about health, financial status, or political leanings) that they never disclosed.[151] Making inferential transparency a regulatory requirement would help bridge this knowledge gap. Inferential transparency includes disclosing categories of inferences and their sources, intended use, and potential effects on people.[152]

Second, transparency should extend to the purposes and contexts of data uses. Existing requirements to state "purposes of processing" are often framed in vague or broad terms, such as "product improvement" or "service personalization."[153] These formulations are insufficient to allow people to estimate the risk that a data use might have. Reducing structural uncertainty through transparency requires that data controllers provide concrete illustrations of how personal data is operationalized in contexts that affect people. This is particularly relevant for data-driven decision-making systems—for example, in determining eligibility for financial services or tailoring content in recommender algorithms.[154] This form of contextual transparency would improve people's ability to anticipate some real-world implications of the forms of data collection they are asked to agree to.

---

[150] Sandra Wachter & Brent Mittelstadt, *A Right to Reasonable Inferences: Re-Thinking Data Protection Law in the Age of Big Data and AI*, 2019 COLUM. BUS. L. REV. 494, 499–507 (2019) (introducing a "right to reasonable inferences," noting GDPR fails to require controllers to disclose what inferences they generate or how they are used); *see also* Data Protection Working Party, *Guidelines on Automated Individual Decision-Making and Profiling for the Purposes of Regulation 2016/679*, at 31, 17/EN, WP251rev.01 (Feb. 6, 2018) (recommending that controllers provide "meaningful information" about how profiles are used and why they are relevant to decisions).

[151] Alicia Solow-Niederman, *Information Privacy and the Inference Economy*, 117 NW. U. L. REV. 357, 361 (2022); Cofone & Robertson, *supra* note 127.

[152] Wachter & Mittelstadt, *supra* note 150, at 543–44.

[153] van de Waerdt, *supra* note 147 (noting that GDPR's stated purposes are routinely framed too generically to empower consent or understanding); Grundler et al., *supra* note 147, at 72, 74–75, 81 (privacy policy language often refers to imprecise data categories and vague processing purposes).

[154] *See* Margot E. Kaminski, *Binary Governance: Lessons from the GDPR's Approach to Algorithmic Accountability*, 92 S. CAL. L. REV. 1529,

Third, transparency should include information about downstream data flows.[155] Downstream flows include data sharing not only with immediate service providers, but also with data brokers, advertisers, and analytics platform providers, many of whom may further disseminate the data.[156] People lack visibility into how their data travels through these systems of third-party processors and controllers. To reduce uncertainty about future data uses and exposures, data controllers should disclose not only the categories of third parties involved, but also the logic of data sharing arrangements, including whether those third parties engage in profiling or automated decision-making. One possibility is a layered transparency model: a high-level explanation accessible to lay users followed by a detailed, machine-readable presentation of data flows meant for regulators, auditors, and researchers.[157]

Fourth, regulators should encourage dynamic transparency: updating disclosures near real time when the nature or purpose of data use changes in ways that can materially affect initial decisions.[158] Static privacy notices, even if initially detailed, lose their relevance as data practices evolve. Dynamic transparency mechanisms, such as privacy dashboards, interactive notifications, and "data use alerts," help reduce this temporal gap.[159] Reducing the gap is especially helpful in contexts where people are re-exposed to risks they could not have predicted at the time of initial data collection.[160] This involves setting different types of transparency for

---

1588–89 (2019) (arguing that algorithmic accountability requires disclosing how data is used in decision systems and not just listing categories).

[155] Jeremy Berkowitz, Michael Mangold & Stephen Sharon, *Data Flow Maps–Increasing Data Processing Transparency and Privacy Compliance in the Enterprise*, 73 WASH. & LEE L. REV. ONLINE 802, 815–16 (2017) (proposing disclosing the structure of data flows within and outside an organization), https://scholarlycommons.law.wlu.edu/wlulr-online/vol73/iss2/11/ [https://perma.cc/8ARK-UJKU].

[156] *Id*.

[157] Kaminski, *supra* note 154, at 1535–36 (explaining the GDPR's transparency as layered: simplified disclosures to individuals and more technical ones for regulators).

[158] *See* CTR. FOR INFO. POL'Y LEADERSHIP, RECOMMENDATIONS ON TRANSPARENCY, CONSENT AND LEGITIMATE INTEREST UNDER THE GDPR 9–11 (2017) (advocating for embedding transparency into user experience with real-time updates and interactive dashboards).

[159] *Id.*

[160] *See* Florian Schaub, Rebecca Balebako, Adam L. Durity & Lorrie Faith Cranor, *A Design Space For Effective Privacy Notices*, *in* 11TH

different audiences. Although dynamic transparency could be overwhelming for individual users, it can be helpful for other audiences under a layered transparency model.

These enhanced forms of transparency move beyond the disclosures required by privacy law.[161] They aim to make the risks of data collection more legible in behavioral terms, enabling people to better align their choices with their preferences. In doing so, such measures could reduce the structural uncertainty, driven by information asymmetries, that produces seemingly paradoxical behavior; not by attempting to change people, but by changing the decision environment.

### B. FUNCTIONAL PRIVACY POLICIES

The findings of this Article, which suggest that privacy decisions are made under conditions of structural uncertainty, reinforce the need to reassess the content, design, and presentation of privacy policies. People seek flexibility because they make privacy decisions under uncertain risk. However, data controllers can make that risk intentionally uncertain by obfuscating it with uninformative privacy policies and other mechanisms such as dark patterns.[162] As a result, people make decisions based on more uncertainty than is necessary. Regulatory measures should counter this practice.

One way to address structural uncertainty is to target the language and function of privacy policies.[163] Rather than treating privacy policies as instruments merely designed to enable data collection and allocate liability, regulators should treat them as tools for facilitating decision-making under risk.

---

SYMPOSIUM ON USABLE PRIVACY & SECURITY 1 (2015); Florian Schaub, Rebecca Balebako & Lorrie Faith Cranor, *Designing Effective Privacy Notices And Controls*, *in* 21.3 IEEE INTERNET COMPUTING 70 (2017).

[161] *See* Wachter & Mittelstadt, *supra* note 150, at 502–05.

[162] Jamie Luguri & Lior Jacob Strahilevitz, *Shining a Light on Dark Patterns*, 13 J. LEGAL ANALYSIS 43, 52–54 (2021) (finding how dark patterns manipulate users into undesired privacy choices by increasing perceived difficulty and uncertainty in opting out).

[163] Susanne Barth, Dan Ionita & Pieter Hartel, *Understanding Online Privacy—A Systematic Review of Privacy Visualizations and Privacy by Design Guidelines*, 55 ACM COMPUT. SURVS. 1, 18 (2023), https://doi.org/10.1145/3502288. *See generally* Joel R. Reidenberg, Jaspreet Bhatia, Travis D. Breaux & Thomas B. Norton, *Ambiguity in Privacy Policies and the Impact of Regulation*, 45 J. LEGAL STUD. S163 (2016). *Cf.* Adam S. Chilton & Omri Ben-Shahar, *Simplification of Privacy Disclosures: An Experimental Test* (U. Chi. L. Sch., Working Paper No. 737, 2016).

Most privacy policies are difficult to read and uninformative when read. If an average person reads all privacy policies offered to them during the year in their entirety, it would take them 201 hours per year.[164] Seventy percent of people consider privacy policies difficult to understand and often do not read them for that reason.[165] Most people lack the capacity to understand intricate yet vague privacy policies if they do read them, leading to decisions that may appear paradoxical.[166]

People have been shown to value privacy when it is presented in an understandable way.[167] At a general level, privacy policies should improve their intelligibility.[168] A digested summary at the top stating the most relevant elements saliently, with detailed disclosures below, can increase transparency, as the experimental literature on salient notices suggests.[169] These summaries can include a set of predefined items that match enhanced transparency obligations; for example, whether the company may share the user's information with third parties, specific authorized purposes for processing, and whether the information is deleted once the user removes it from the system.

---

[164] Aleecia M. McDonald & Lorrie Faith Cranor, *The Cost of Reading Privacy Policies*, 4 I/S: J.L. & POL'Y FOR INFO. SOC'Y 543, 565–66 (2008) (demonstrating that reading privacy policies would require an estimated 201 hours per year, making them impractical and unreadable for most users).

[165] Jasmin Kaur, Rozita Dara & Ritu Chaturvedi, *A Semantic-Based Approach To Reduce The Reading Time Of Privacy Policies*, 19TH ANNUAL INT'L CONF. PRIVACY, SEC. & TRUST 1 (2022); JOSEPH TUROW, L. FELDMAN & K. MELTZER, OPEN TO EXPLOITATION: AMERICA'S SHOPPERS ONLINE AND OFFLINE (Annenberg Pub. Pol'y Ctr. U. Pa. 2005); George R. Milne & Mary J. Culnan, *Strategies for Reducing Online Privacy Risks: Why Consumers Read (or Don't Read) Online Privacy Notices*, 18 J. INTERACTIVE MKTG. 15, 15–29 (2004); Reidenberg et al., *supra* note 138, at 40–41, 83–87 (showing that most users misunderstand or ignore privacy policies because they are too complex or misleading); PATRICK GAGE KELLEY, LUCIAN CESCA, JOANNA BRESEE & LORRIE FAITH CRANOR, STANDARDIZING PRIVACY NOTICES: AN ONLINE STUDY OF THE NUTRITION LABEL APPROACH (2010).

[166] Reidenberg et al., *supra* note 138, at 40–41, 83–87.

[167] Gerber et al., *supra* note 33, at 252, 255; Mourey & Waldman, *supra* note 38, at 162; John et al., *supra* note 34, at 868; *see also* Adam Shostack & Paul Syverson, *What Price Privacy? (and Why Identity Theft is About Neither Identity nor Theft)*, *in* ECONOMICS OF INFORMATION SECURITY 129 (L. Jean Camp & Stephen Lewis eds., 2004); Ebert et al., *supra* note 32, at 11; *cf.* Groom & Calo, *supra* note 32, at 4.

[168] *See, e.g.*, Preet Sanghavi, Raj Ghamsani, Rishi Parekh, Ritik Mota & Deepika Dongre, *Simplifying Privacy Agreements Using Machine Reading Comprehension and Open Domain*, 6th INT'L CONF. ON COMPUT., COMMC'N., CONTROL & AUTOMATION 1 (2022); Nazila Gol Mohammadi, Julia Papmus & Maritta Heisel, *Pattern-based Incorporation of Privacy Preferences into Privacy Policies*, PROC. 24TH EUR. CONF. PATTERN LANGUAGES OF PROGRAMS 1 (2019); PATRICK GAGE KELLEY, JOANNA BRESEE, LORRIE FAITH CRANOR & ROBERT W. REEDER, A "NUTRITION LABEL" FOR PRIVACY 4 (2009).

One could require standardization in the presentation of key information, such as data retention periods, categories of third-party sharing, whether profiling is used, and use of automated decision-making.[170] This would allow people to more easily compare privacy practices across services. Similarly to how standardized nutrition labels improve consumer awareness of dietary risks, standardized privacy labels could reduce uncertainty under time constraints.[171] The introduction of standardized formats—mandating the disclosure of calories, fat content, sugar, and other metrics in a clear, comparable structure—significantly improved consumers' ability to assess health risks at a glance.[172] This standardization benefited not only those who read labels, but also those making decisions with limited time. The success of that model is not dependent on complete consumer understanding: it rests on reducing cognitive friction and increasing the salience of key information. Privacy labels could do the same. A clear, consistent format that highlights core elements would enable entities to compare policies.[173]

---

[169] KELLEY ET AL., *supra* note 165, at 1574–76 (finding that traditional privacy policies are largely ignored, but more salient formats increased engagement and comprehension); *see also* Lorrie Faith Cranor, *Necessary But Not Sufficient: Standardized Mechanisms For Privacy Notice and Choice*, COLO. TECH. L.J. 273, 287 (2012) (advocating for layered privacy notices, showing users engage more with upfront, digestible privacy summaries than with full-text policies); Ebert, *supra* note 32 (showing concise privacy notices are recalled better when they are made salient); Vanessa Bracamonte, Seira Hidano, Welderufael B. Tesfay & Shinsaku Kiyomato, *Evaluating Privacy Policy Summarization: An Experimental Study Among Japanese Users*, 5th INT'L CONF. INFO. SYS. SEC. & PRIV. 370, 376 (2019).

[170] *See* Barth et al., *supra* note 163; Zohar Efroni, Jakob Metzger, Lena Mischau & Marie Schirmbeck, *Privacy Icons: A Risk-Based Approach to Visualisation of Data Processing*, 5 EUR. DATA PROT. L. REV. 352, 358 (2019); *see also* Sanghavi et al., *supra* note 168, at 3–6 (discussing clear structure and concise data practices explanations).

[171] Kelley T. Watson & Paul G. Barash, *The New Food and Drug Administration Drug Package Insert: Implications for Patient Safety and Clinical Care*, 108 ANESTHESIA & ANALGESIA 211, 211 (2009), https://doi.org/10.1213/ane.0b013e31818c1b27; Patrick Gage Kelley, Lucian Cesca, Joanna Bresee & Lorrie F. Cranor, *Standardizing Privacy Notices: An Online Study of the Nutrition Label Approach*, *in* PROCEEDINGS OF THE SIGCHI CONFERENCE ON HUMAN FACTORS IN COMPUTING SYSTEMS 1573, 1580–81 (2010), https://doi.org/10.1145/1753326.1753561 (showing that standardized privacy labels help users more quickly and accurately assess company practices, especially under time pressure); FED. TRADE COMM'N, PROTECTING CONSUMER PRIVACY IN AN ERA OF RAPID CHANGE: RECOMMENDATIONS FOR BUSINESSES AND POLICYMAKERS 61–62 (2012) (recommending short-form, standardized privacy notices, comparing them directly to nutrition labels as a model for improving understanding).

Making privacy policies functional at reducing structural uncertainty requires a fuller conception of their completeness. Many privacy policies omit significant data collection practices, particularly those involving opaque forms of tracking.[174] Completeness should be seen as more than full descriptions of data collection, including complete descriptions of forms of data processing, data security measures, and data sharing specifics. The experimental literature reviewed points to the relevance of context in privacy decisions.[175] Considering the contextual differences that are relevant to people can reduce uncertainty over risk.[176]

Yet completeness is insufficient. Because most people lack either the time or capacity to process long and complex documents (or both), reforms must also address the *accessibility* and *salience* of privacy policies. Experimental evidence, including the findings presented in this Article, suggests that privacy decisions are shaped not only by the content of information but by how and when information is presented. [177] Much relevant information, including privacy policies, sits at the periphery of user interfaces and is easily ignored.[178] A low-cost reform could require that uncertainty-reducing information appears prominently, for example during account creation. While this change would not increase comprehension on its own, it would improve the status quo if paired with salience mechanisms. These measures would reduce uncertainty if paired with visual indicators and layered disclosures.

---

[172] *See* Yan Shvartzshnaider, *Privacy Inserts*, BALKINIZATION (Dec. 7, 2024), https://balkin.blogspot.com/2024/12/privacy-inserts.html [https://perma.cc/NCE3-3JFH].

[173] *Id.*

[174] Julissa Milligan, Sarah Scheffler, Andrew Sellars, Trishita Tiwari, Ari Trachtenberg & Mayank Varia, *Case Study: Disclosure of Indirect Device Fingerprinting in Privacy Policies*, in PRIVACY TECHNOLOGIES & POLICY 175, 178–83 (2021) (finding sites using browser fingerprinting rarely disclosed these practices in their privacy policies).

[175] Tobias Dienlin, Miriam J. Metzger & Seungwoo Lee, *A Longitudinal Analysis of the Privacy Paradox*, 25 NEW MEDIA SOC'y 1043, 1058–59 (2023) (discussing the importance of context in terms of the circumstances under which users are making privacy decisions).

[176] NISSENBAUM, *supra* note 35 at 149–51 (showing that privacy expectations vary based on context, and that disclosures should align with those expectations.)

[177] *See* Tsai et al., *supra* note 31.

[178] *See, e.g.*, Kelley et al., *supra* note 165, at 1574–76.

Incorporating visual indicators means using icons, infographics, and data-flow diagrams to summarize key disclosures (e.g., purposes, data categories, recipients, retention, legal bases, rights) in the summary layer of a policy and at points of collection.[179] These visuals are meant to supplement the governing text and be cross-referenced to specific sections.[180] Layered disclosures involve giving privacy policies a tiered structure: high-level, plain-language disclosures first followed by more comprehensive, technical disclosures in subordinate sections, permitting variable depth of review and allowing people to access information at their preferred level of granularity.[181]

Finally, reducing structural uncertainty is incompatible with interfaces that manufacture or exploit that uncertainty.[182] So enhanced transparency should involve prohibiting and penalizing deceptive and manipulative practices within and beyond privacy policies.[183] Deceptive design (such as salience distortion, asymmetric friction, equivocal wording, or visual shrouding) widens the gap between what people need to know to evaluate consequences and what they actually perceive at the moment of choice.[184] It does so by manipulating the decision environment that transparency is meant to clarify. When interfaces inflate variance in people's understanding of potential harm, decisions cannot operate as informed responses to risk. Banning designs that increase uncertainty about consequences (or that impede access to mitigation controls) aligns with the regulatory objective of enhanced transparency and of requiring privacy policies.

---

[179] Barth et al., *supra* note 163, at 2; Aikaterini Soumelidou & Aggeliki Tsohou, *Effects of Privacy Policy Visualization on Users' Information Privacy Awareness Level: The Case of Instagram*, 33 INFO. TECH. & PEOPLE 502, 505–06 (2020); Daniel Reinhardt, Johannes Borchard & Jörn Hurtienne, *Visual Interactive Privacy Policy: The Better Choice?*, *in* CHI CONFERENCE ON HUMAN FACTORS IN COMPUTING SYSTEMS 1 (2021); Ioannis Paspatis, Aggeliki Tsohou & Spyros Kokolakis, *AppAware: A Policy Visualization Model for Mobile Applications*, 28 INFO. & COMP. SEC. 116 (2020).

[180] Efroni, *supra* note 171, at 359.

[181] Armin Gerl & Bianca Meier, *Privacy in the Future of Integrated Health Care Services—Are Privacy Languages the Key?*, *in* INTERNATIONAL CONFERENCE ON WIRELESS AND MOBILE COMPUTING, NETWORKING & COMMUNICATIONS 312 (2019); JENS LEICHT, ARMIN GERL & MARITTA HEISEL, TECHNICAL REPORT ON THE EXTENSION OF THE LAYERED PRIVACY LANGUAGE (2021).

[182] Daniel Susser, Beate Roessler & Helen Nissenbaum, *Technology, Autonomy, and Manipulation*, 8 INTERNET POL'Y REV. 1, 4 (2019); Kirsten Martin, *Manipulation, Privacy, and Choice*, 23 N.C. J.L. & TECH. 452, 458, 502 (2022).

[183] Waldman, *supra* note 21, at 107; Mark Leiser & Cristiana Santos, *Dark Patterns, Enforcement, and the Emerging Digital Design Acquis: Manipulation Beneath the Interface*, 15 EUR. J.L. & TECH. 1, 16–17 (2024); Luguri & Strahilevitz, *supra* note 162, at 82–83; Martin Brenncke, *Regulating Dark Patterns*, 14 NOTRE DAME J. INT'L & COMP. L. 41, 43 (2023).

### C. THE RIGHT TO BE FORGOTTEN AS A FLEXIBILITY MECHANISM

Debates about the right to be forgotten (RTBF) have often centered on its limits, particularly its tension with freedom of expression and archival interests.[185] However, the right also has a systemic effect on online interactions.[186] The right's role as a *flexibility mechanism* has received less attention.[187]

In its GDPR formulation, the RTBF allows people to request the erasure or de-linking of their personal data when certain conditions are met, such as when data is no longer necessary for the purposes for which it was collected, the individual withdraws consent, or the processing is unlawful.[188]

---

[184] *See generally* Matthew B. Kugler, Lior Strahilevitz, Marshini Chetty, Chirag Mahapatra & Yaretzi Ulloa, *Can Consumers Protect Themselves Against Privacy Dark Patterns?*, 23 U. N.H. L. Rev. 243, 246–47 (2025).

[185] *See, e.g.*, W. Gregory Voss & Céline Castets-Renard, *Proposal for an International Taxonomy on the Various Forms of the "Right to Be Forgotten": A Study on the Convergence of Norms*, 14 COLO. TECH. L.J. 281, 292–293 (2016), https://scholar.law.colorado.edu/ctlj/vol14/iss2/6/ [https://perma.cc/3XLQ-KW4M]; Eloïse Gratton & Jules Polonetsky, *Droit à l'oubli: Canadian Perspective on the Global 'Right to Be Forgotten' Debate*, 15 COLO. TECH. L.J. 337, 343 (2017), https://ctlj.colorado.edu/wp-content/uploads/2017/09/4-GrattonPolo.pdf [https://perma.cc/ZGH4-AGUM]; Edward Lee, *The Right to Be Forgotten v. Free Speech*, 12 I/S J. L. POL'Y. FOR INFO. SOC'Y 85, 92 (2015); Antoon De Baets, *A Historian's View on the Right to Be Forgotten*, 30 INT'L REV. L. COMPTS. & TECH. 57, 58 (2016), https://doi.org/10.1080/13600869.2015.1125155 [https://sci-hub.box/https://doi.org/10.1080/13600869.2015.1125155].

[186] Christopher S. Yoo, *An Economic Analysis of the Right to Be Forgotten*, U. PA. INST. FOR L. & ECON., Rsch. Paper No. 22-25 (2023).

[187] *See generally* Theo Bertram, Elie Bursztein, Stephanie Caro, Hubert Chao, Rutledge Chin Feman, Peter Fleischer, Albin Gustafsson, Jess Hemerly, Chris Hibbert, Luca Invernizzi, Lanah Kammourieh Donnelly, Jason Ketover, Jay Laefer, Paul Nicholas, Yuan Niu, Harjinder Obhi, David Price, Andrew Strait, Kurt Thomas & Al Verney, *Five Years of the Right to Be Forgotten*, *in* INTERNATIONAL CONFERENCE ON WIRELESS AND MOBILE COMPUTING, NETWORKING, AND COMMUNICATIONS 959 (2019), https://doi.org/10.1145/3319535.3354208 (providing an overview of RTBF requests).

In both forms, the RTBF is best understood not only as a means of enforcing data minimization or protecting reputational interests, but also as a mechanism for managing uncertainty in privacy decision-making. The right enables people to revise earlier data collection decisions made under conditions of uncertainty: it provides them with flexibility when deciding whether to disclose personal information by allowing them to reverse decisions.[189] Most RTBF petitions in the EU are not about false information; they target truthful, lawfully published facts that people no longer want to have define them.[190] That is why anchoring the RTBF solely in information accuracy is a poor fit: the issue it addresses is the continued prominence of past facts in name-search results.[191]

This framing is supported by EU institutional commentary. The European Commission stated in its proposal for the GDPR that a "reinforced 'right to be forgotten' will help people better manage data protection risks online."[192] Similarly, the Article 29 Working Party emphasized that the right helps empower people to request the deletion of their personal data, thus offering a means to exercise control over their own digital identity.[193] This understanding aligns

---

[188] Regulation (EU) 2016/679 of the European Parliament and of the Council of 27 Apr. 2016, on the Protection of Natural Persons with Regard to the Processing of Personal Data and on the Free Movement of Such Data, and Repealing Directive 95/46/EC (General Data Protection Regulation), art. 17(1)(a), (b), (d), 2016 O.J. (L 119) 1, 43 (EU).

[189] *See generally* VIKTOR MAYER-SCHÖNBERGER, DELETE: THE VIRTUE OF FORGETTING IN THE DIGITAL AGE (2009) (providing a general account of the right consistent with this view), https://doi.org/10.1515/9781400838455.

[190] Teresa Scassa, *A Little Knowledge Is a Dangerous Thing?: Information Asymmetries and the Right to Be Forgotten*, *in* THE RIGHT TO BE FORGOTTEN: A CANADIAN AND COMPARATIVE PERSPECTIVE (Ignacio Cofone ed., 2020), 26, 27–39.

[191] *Id.*

[192] European Commission MEMO/12/41, Data Protection Reform: Frequently Asked Questions (Jan. 25, 2012), https://ec.europa.eu/commission/presscorner/detail/en/memo_12_41 [https://perma.cc/K939-Z35D].

[193] Viviane Reding, Vice-President of the Eur. Comm'n, EU Just. Comm'n, Speech at the Munich Digital, Life, Design Innovation Conference: The EU Data Protection Reform 2012: Making Europe the Standard Setter for Modern Data Protection Rules in the Digital Age (Jan. 22, 2012) ("If an individual no longer wants his personal data to be processed or stored by a data controller, and if there is no legitimate reason

with the account in this Article: privacy choices are often made under uncertainty. In such cases, the ability to revisit those decisions is a policy response to information asymmetry.

Early case law from the European Court of Justice (ECJ) establishing the right partially confirms this view. In *Google Spain*,[194] the court recognized that people have the right to request the delisting of search engine results that link to personal data that is "inadequate, irrelevant or no longer relevant." The Court grounded this right in the GDPR's predecessor (the Data Protection Directive),[195] and the fundamental rights to privacy and data protection under the Charter of Fundamental Rights of the European Union.[196] The *Google Spain* ruling highlighted that people should not be permanently bound by the consequences of past data collection when those no longer serve a legitimate public interest, supporting the perspective that the RTBF is a flexibility mechanism.

Subsequent case law that refined and contextualized the scope of the RTBF aligns with the flexibility account too. In *GC and Others v. CNIL*, when addressing the geographic scope of delisting, the Court clarified that the RTBF must be balanced against the rights to freedom of expression and access to information.[197] The ECJ reiterated that data protection rights are not absolute and must be weighed against competing fundamental rights.[198] Yet even within that balancing framework, the court affirmed the legitimacy of people seeking to revise their online presence, particularly when the continued availability of certain search results causes disproportionate harm relative to their public value.[199]

---

for keeping it, the data should be removed from their system."); *see also* Article 29 Data Protection Working Party, *Opinion 01/2014 on the Application of Necessity and Proportionality and Data Protection in Law Enforcement* 536/14/EN WP 211 7–9, 21 (Feb. 27, 2014); *see also Proposal for a Regulation of the European Parliament and of the Council on the Protection of Individuals with Regard to the Processing of Personal Data and on the Free Movement of Such Data (General Data Protection Regulation)*, at 25–26, COM (2012) 11 final (Jan. 25, 2012).

[194] Google Spain SL v. Agencia Española de Protección de Datos (AEPD), Case C-131/12, ECLI:EU:C:2014:317 (May 13, 2014).

[195] Council Directive 95/46/EC, art. 12(b), 1995 O.J. (L 281) (EU).

[196] Charter of Fundamental Rights of the European Union No. 2012/C 326/02, arts. 7, 8, 2012 O.J. (C 326/391).

[197] GC and Others v. Commission nationale de l'informatique et des libertés (CNIL), Case C-136/17, EU:C:2019:773 at para 89 (Sept. 24, 2019).

[198] See *infra* notes 188–90 and accompanying text.

[199] *See* Sam Wrigley & Anne Klinefelter, *Google LLC v. CNIL: The Location-Based Limits of the EU Right to Erasure and Lessons for U.S. Privacy Law*, 22 N.C. J.L. & TECH. 681, 693–96 (2021),

From a regulatory design perspective, the RTBF's additional flexibility operates as a counterweight to the structural uncertainty people face over data practices in the information economy. People often consent to data processing under bundled terms with limited understanding of the long-term consequences of data collection.[200] Once data enters the system, it can persist indefinitely—and it can be repurposed, recombined, and recontextualized in ways the individual could not have anticipated.[201] The RTBF supplies a corrective mechanism: a structured right to exit or revise one's participation in that system. In behavioral terms, it transforms what might otherwise be an irrevocable choice into a revisable one. That is how the RTBF helps individuals to curate their digital presence in accordance with their values and life circumstances.[202]

This framing is related to the arguments that the RTBF enables freedom to change as an element of self-development,[203] and that control over one's personal information is central to information privacy.[204] The structural uncertainty account presented here indicates that the RTBF has value beyond a deontological, rights-

---

https://scholarship.law.unc.edu/cgi/viewcontent.cgi?article=1428&context=ncjolt [https://perma.cc/N3WT-G6HM].

[200] *See* Strandburg, *supra* note 71; Cofone & Robertson, *supra* note 127.

[201] Solow-Niederman, *supra* note 151, at 379–84, 411–13; Daniel J. Solove, *Privacy Self-Management and the Consent Dilemma*, 126 HARV. L. REV. 1880, 1889–91 (2013). *See also* Eugenia Politou, Efthimios Alepis & Constantinos Patsakis, *Forgetting Personal Data and Revoking Consent Under the GDPR: Challenges and Proposed Solutions*, 4 J. CYBERSECURITY 1, 2–4 (2018).

[202] This relates more broadly to the misalignment between mutable human identity and immutable data traces. As preferences, reputations, and social contexts evolve, the continued availability of outdated or irrelevant personal data can distort the individual's autonomy. *See* Chanhee Kwak, Junyeong Lee & Heeseok Lee, *Could You Ever Forget Me? Why People Want to Be Forgotten Online*, 179 J. BUS. ETHICS 25, 26–28 (2022). *See generally* MAYER-SCHÖNBERGER, *supra* note 190.

[203] Jean-François Blanchette & Deborah G. Johnson, *Data Retention and the Panoptic Society: The Social Benefits of Forgetfulness*, 18 INFO. SOC'Y 33, 35 (2002); Chris Conley, *The Right to Delete*, 2010 ASS'N ADVANCEMENT A.I. SPRING SYMP.: INTELLIGENT INFO. PRIV. MGMT. 53, 53–54; Franz Werro, *The Right to Inform v. The Right to be Forgotten: A Transatlantic Clash*, *in* HAFTUNGSRECHT IM DRITTEN MILLENNIUM [LIABILITY IN THE THIRD MILLENNIUM] 285, 285–87 (Aurelia Colombi Ciacchi, Christine Godt, Peter Rott & Lesley Jane Smith eds., 2009); s*ee also* Politou et al., *supra* note 202, at 9–10.

[204] Data Protection Working Party, *Opinion 15/2011 on the Definition of Consent*, WP187, Doc. No. 01197/11/EN, at 5–6, 8–9, 33 (2011).

based perspective because flexibility enables self-development, an idea supported by philosophical accounts of digital rights.[205] In response to critics of the RTBF who note its potential chilling effects on public discourse, archival integrity, and freedom of expression,[206] ECJ case law incorporated proportionality and contextual balancing tests.[207] The GDPR excludes the application of the RTBF where processing is necessary for exercising the right of freedom of expression, complying with legal obligations, or performing tasks in the public interest.[208] These safeguards, which make the RTBF conditional instead of absolute, fit with a flexibility-providing role.

In other words, a regulatory structure that includes the RTBF is aligned with empirical insights about how people disclose personal information under uncertainty.[209] By enabling people to reverse or revise past data collection, the RTBF embeds some level of flexibility into the architecture of data protection law.

This view also has implications for reform. Generative AI systems intensify profile compilation, pulling scattered items into a single profile (and often resurfacing low-value facts). URL delisting only does not suffice because models can paraphrase without linking.[210] The RTBF, seen as a flexibility mechanism, should extend from search engines to large language model outputs for name prompts since the reason to apply it to search engines is applicable to them as well.

---

[205] Lowry Pressly, *The Right to Be Forgotten and the Value of an Open Future*, 135 ETHICS 65, 72–75 (2024) (arguing that, by enabling the take-down of personal data, the RTBF supports self-development and people's evolving self-concept, as who we are (and what we consented to share) at one time may no longer reflect who we become).

[206] *See, e.g.*, Robert C. Post, *Data Privacy and Dignitary Privacy: Google Spain, the Right To Be Forgotten, and the Construction of the Public Sphere*, 67 DUKE L.J. 981, 1008–09 (2018).

[207] *See generally* Silvia De Conca, *GC et al. v. CNIL: Balancing the Right to Be Forgotten with the Freedom of Information, the Duties of a Search Engine Operator*, 5 EUR. DATA PROT. L. REV. 561, 563 (2019); Kyu Ho Youm & Ahran Park, *The Right to Be Forgotten: Google Spain as a Benchmark for Free Speech Versus Privacy?*, 24 CHI. J. INT'L L. 167, 173 (2023); Stefan Kulk & Frederik Zuiderveen Borgesius, *Freedom of Expression and 'Right to Be Forgotten' Cases in the Netherlands After Google Spain*, 1 EUR. DATA PROT. L. REV. 113, 122 (2015); Theo Bertram, Elie Bursztein, Stephanie Caro, Hubert Chao, Rutledge C. Feman, Peter Fleischer, Albin Gustafsson, Jess Hemerly, Chris Hibbert, Luca Invernizzi, Lanah K. Donnelly, Jason Ketover, Jay Laefer, Paul Nicholas, Yuan Niu, Harjinder Obhi, David Price, Andrew Strait, Kurt Thomas & Al Verney, *Five Years of the Right to Be Forgotten*, *in* PROCEEDINGS OF THE CONFERENCE ON COMPUTER AND COMMUNICATIONS SECURITY 959 (2019).

[208] General Data Protection Regulation 2016/679, art. 17(3), 2018 O.J. (L 119) (EU).

[209] *See generally* Acquisti et al., *supra* note 48, at 510–13; Yoo, *supra* note 187, at 14–17.

## CONCLUSION

This Article shows that, because privacy decisions occur under structural uncertainty, behavior that might appear inconsistent is in fact a rational response to risk.

This Article introduces a novel experimental test that, building on prior work on information asymmetries, examines the mechanisms behind the so-called privacy paradox. The study isolates discounting mechanisms to identify what drives privacy behavior and offers a novel explanation: people's behavior reflects responses to uncertain risk. That is, people when acting as data subjects do not have different time preferences than when they act as standard consumers; they just respond to a different context.

As a consequence, it finds that attributing privacy choices entirely to cognitive biases overlooks the role of uncertainty in decision-making. Assuming that people behave inconsistently when they express they value their privacy while disclosing information rests on the assumption that the risks associated with data collection are known and quantifiable. But that assumption does not hold in most real-world contexts. People deciding whether to share personal information (or agree to its collection, use, or disclosure) face profound uncertainty about how their data will be used, who will access it, how long it will persist and, most importantly, what downstream consequences may result. There is nothing contradictory about the behavior of people who value their privacy and agree to data practices in a context of uncertain future harms. The privacy paradox is a misnomer.

This reframing carries significant regulatory implications. This structural uncertainty account sees people as operating under constraints in a decision environment—and calls for addressing those constraints. As privacy behavior reflects responses to structural uncertainty rather than failures of self-control, the policy tools usually proposed to correct the latter behavior—such as default settings or behavioral nudges—may be misdirected. These tools are built on the assumption that people need help managing their impulses, whereas people may need more transparency over risks and flexibility. In this sense, the policy recommendations that flow from a structural uncertainty account reflect the demands voiced by consumer advocates and policymakers.

---

[210] Dawen Zhang, Pamela Finckenberg-Broman, Thong Hoang, Shidong Pan, Zhenchang Xing, Mark Staples & Xiwei Xu, *Right to Be Forgotten in the Era of Large Language Models: Implications, Challenges, and Solutions*, 5 AI & ETHICS 2445, 2450 (2025).

The regulatory response to this account lies in changing the structure of the decision environment to reduce uncertainty over risks. Transparency obligations should be enhanced by requiring not only disclosures of data collection, but also by making those disclosures specific and context-sensitive. Consent-supporting mechanisms, such as privacy policies, should be reformed to support uncertainty-reduction. The right to be forgotten is valuable for accurate information because it allows people to revise or retract past data collection as their understanding of risk evolves. Privacy law benefits people beyond reputational interests when it recognizes the need to revisit decisions made under uncertainty.

## APPENDIX A: MATERIALS

### A. SURVEY

Respondents registered on Qualtrics received an invitation from the system to complete a short survey.

The initial encounter with respondents built on language used by Acquisti, John, and Loewenstein,[211] who approached respondents at a shopping mall and asked them to complete a brief survey designed to assess people's attitudes toward spending money. The words "tracked" and "privacy" were avoided to prevent priming respondents.

The initial question was:

*"We are conducting a brief survey designed to assess people's attitudes when spending money. The survey will take you approximately 1 minute. After completing the survey, you will receive a $6 or a $9 gift card.*

- *How much do you normally spend on one coffee?*
- *What is your favorite coffee beverage?*
- *Do you prefer hot coffee or iced coffee?*
- *What is your year of birth?*
- *What is your gender?*
- *Are you Hispanic or Latino, or none of these? [Yes/none]*
- *Choose one or more races that you consider yourself to be:*
- *What is the highest level of education you have completed?*
- *What is your ZIP code?"*

---

[211] Acquisti et al., *supra* note 6, at 260–63.

After completing the survey portion of the study, respondents saw a second question block with just one question, which varied depending on the treatment, as shown below.

**Control**

*"Thank you for completing the survey. You will receive your voucher by email within two weeks.*

*You can choose between a $5 voucher, or a larger $7 voucher if you allow us to put your name and favorite beverage in our website.*

*We will send you an email with more details about the website soon, and remind you of this information. You can either choose or let us know when we email you back.*

- *I choose $5*
- *I choose $7*
- *I'll choose later between $5 and $7*

*To send you the voucher, we will need your name and email*

- *What is your full name?*
- *What is your email address?"*

**Treatment 1**

*"Thank you for completing the survey. You will receive your voucher by email within two weeks.*

*You can choose between a $4 voucher, or a larger $6 voucher if you allow us to put your name and favorite beverage in our website.*

*We will send you an email with more details about the website soon, and remind you of this information. You can either choose now or let us know when we email you back. Just so you know, if you respond now, the vouchers will increase to $5 and $7 respectively.*

- *I choose $5*
- *I choose $7*
- *I'll choose later between $4 and $6*

*To send you the voucher, we will need your name and email*

- *What is your full name?*
- *What is your email address?"*

**Treatment 2**

*"Thank you for completing the survey. You will receive your voucher by email within two weeks.*

*You can choose between a $4 voucher, or a larger $6 voucher if you allow us to put your name and favorite beverage in our website.*

*We will send you an email with more details about the website soon, and remind you of this information. You can either choose now or let us know when we email you back. Just so you know, if you choose to wait and respond to our next email, the vouchers will increase to $5 and $7 respectively.*

- *I choose $4*
- *I choose $6*
- *I'll choose later between $5 and $7*

  *To send you the voucher, we will need your name and email*
- *What is your full name?*
- *What is your email address?"*

### B. VOUCHER CHOICE

A week after receiving the first and second email, respondents received an email giving them the choice between both vouchers if they had not chosen yet, or reminding them of the choice if they had. The email also contained some information about the website.

**Control**

*"Thank you again for completing our survey last week.*

*As a reminder, you can choose between a $6 voucher, or a larger $9 voucher if you allow us to put your name and favorite beverage in our website (your age, gender, ethnicity or postal code will not be posted).*

*The website will just list the names of participants of this survey who chose to be part of it, and what is their favorite beverage. It will not be linked directly to any company website."*

**Treatment 1**

*"Thank you again for completing our survey last week.*

*As a reminder, you can choose between a $4 voucher, or a larger $6 voucher if you allow us to put your name and favorite beverage in our website (your age, gender, ethnicity or postal code will not be posted).*

*The website will just list the names of participants of this survey who chose to be part of it, and what is their favorite beverage. It will not be linked directly to any company website."*

**Treatment 2**

*"Thank you again for completing our survey last week.*

*As a reminder, you can choose between a $5 voucher, or a larger $7 voucher if you allow us to put your name and favorite beverage*

*in our website (your age, gender, ethnicity or postal code will not be posted).*

*The website will just list the names of participants of this survey who chose to be part of it, and what is their favorite beverage. It will not be linked directly to any company website."*

### C. DEBRIEF

After the choice is made, respondents receive a last email with a debrief:

*"You have participated in a study designed to learn about privacy attitudes when spending money conducted by Yale University. The results of this study will be useful to understand how we consumers behave and to design better policies for consumer privacy. We hope this study will benefit you as a consumer.*

*Your participation is extremely valuable for our research. There are no known or anticipated risks associated with this study. It was important to avoid mentioning the role of privacy to avoid biased results.*

*The website's link with people's names will not be shared for commercial purposes. After the study is finished, the website will be taken down. Your information will be anonymized at the end of the study. In the meantime, only the researchers involved in this study and those responsible for research oversight will have access to any information that you provided, and all of your personal information will be held in confidence.*

*If you have any questions or would like to know more about the study, email us at yale.privacy.survey@gmail.com. You can also contact Ignacio Cofone at ignacio.cofone@yale.edu.*

*If you would like to talk with someone other than the researchers to discuss problems or concerns, to discuss situations in the event that a member of the research team is not available, or to discuss your rights as a research participant, you may contact the Yale University Human Subjects Committee, 203-785-4688, human.subjects@yale.edu. Additional information is available at http://your.yale.edu/research-support/human-research/research-participants.*

*You will now receive the voucher that we promised.*

*Thank you for your collaboration. Have a wonderful day."*

The voucher followed the debrief immediately.

## APPENDIX B: TESTS

Three tests below compare sample proportions using a 2-sample z-test. P-values were calculated for two-tailed comparisons and results were compared to $p<0.05$ for significance.

The first test (Test A) examines shifts in flexibility when the treatment made choosing “now” worse than in the control. Sample 1 is the proportion of participants who chose “later” in the control. Sample 2 is the proportion of participants who chose “later” in Treatment 2.

| | **Sample 1** | **Sample 2** | **Difference** |
|---|---|---|---|
| Sample proportion | 0.2184874 | 0.6083333 | 0.3898459 |
| 95% CI (asymptotic) | 0.1442 - 0.2927 | 0.521 - 0.6957 | 0.2649 - 0.5147 |
| z-value | 6.1 | | |
| P-value | <0.0001 | | |

Table 5: result of 2-tailed z-test to compare sample proportions

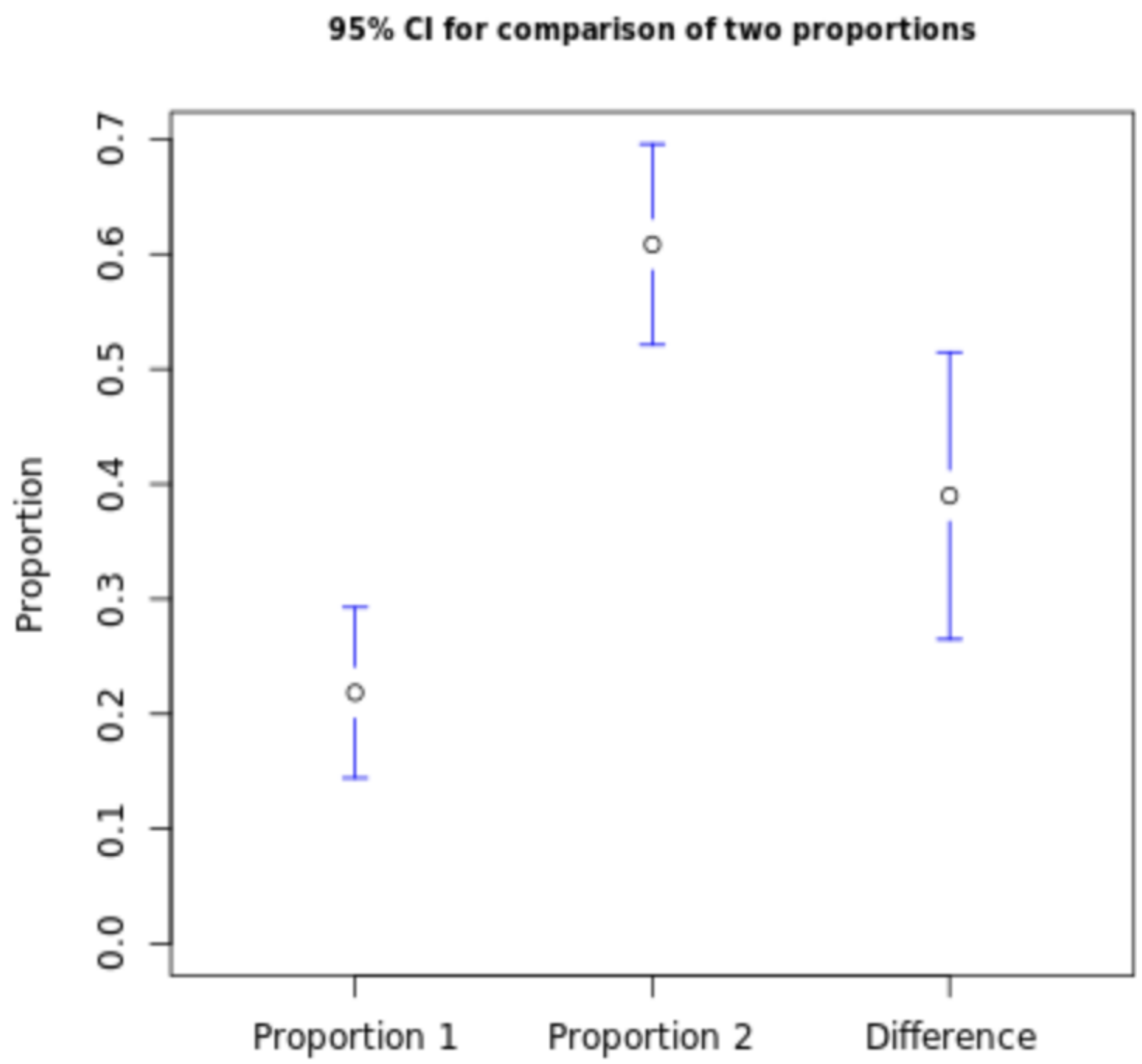

Figure 3: result of 2-tailed z-test to compare sample proportions for Test A The results of Test A are statistically significant. One should reject the hypothesis that proportions of participants choosing flexibility in the control and in Treatment 2 are equal.

The second test (Test B) examines shifts in flexibility when the treatment made choosing "later" worse than in the control. Sample 1 is, as in the last test, the proportion of participants who chose "later" in the control. Sample 2 is the proportion of participants who chose "later" in Treatment 1.

| | **Sample 1** | **Sample 2** | **Difference** |
|---|---|---|---|
| Sample proportion | 0.2184874 | 0.1440678 | 0.0744196 |
| 95% CI (asymptotic) | 0.1442 - 0.2927 | 0.0807 - 0.2074 | -0.0237 - 0.1725 |
| z-value | 1.5 | | |
| P-value | 0.1372 | | |

Table 6: result of 2-tailed z-test to compare sample proportions

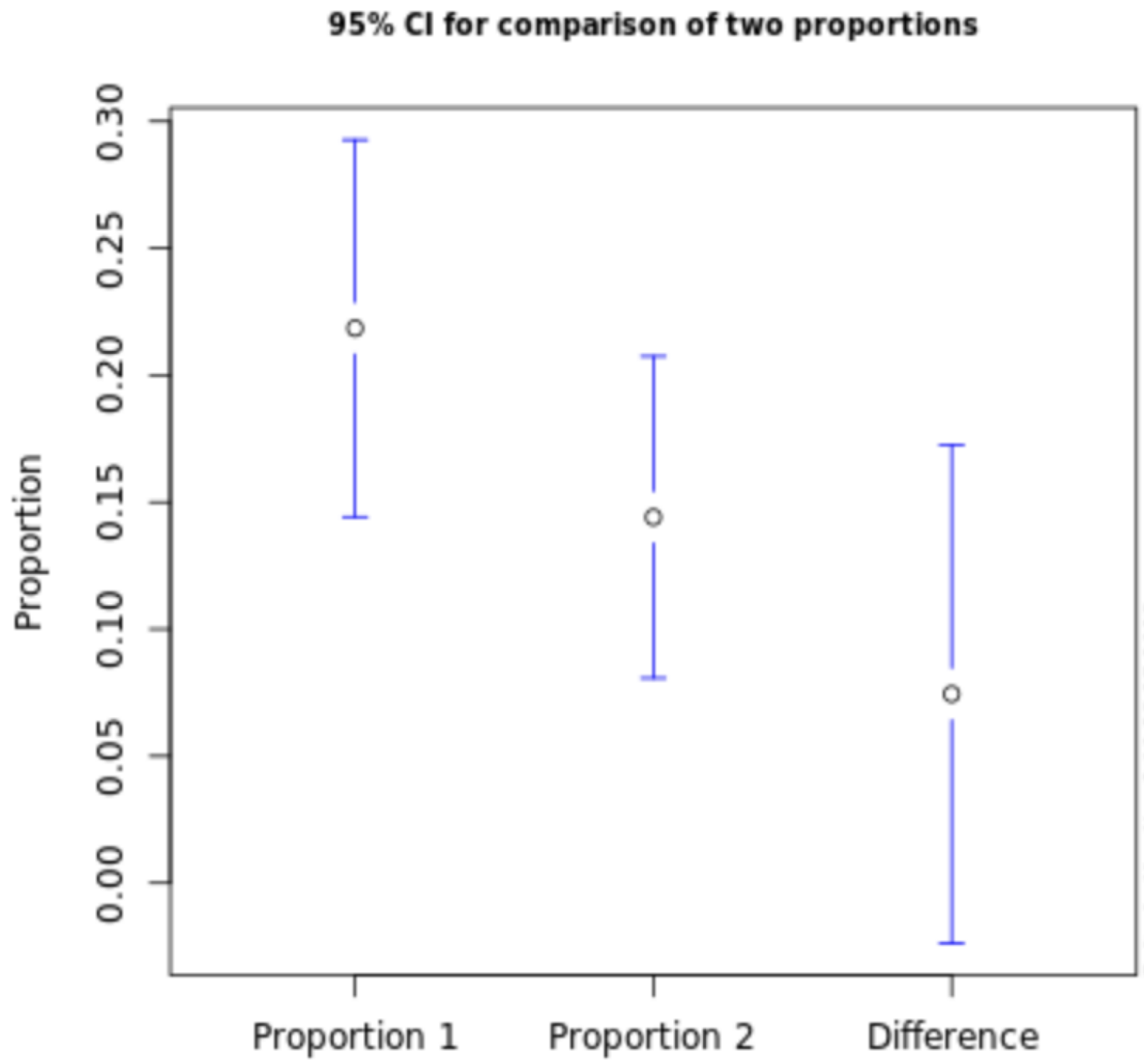

Figure 4: result of 2-tailed z-test to compare sample proportions for Test B

The results of Test B are not statistically significant. One cannot reject the null hypothesis that the proportions of participants choosing flexibility in the control and in Treatment 1 are equal.

The third test (Test C) examines shifts in pre-commitment when the treatment made choosing "now" worse than in the control. Sample 1 is the proportion of participants who chose the privacy voucher in the control. Sample 2 is the proportion of participants who chose the privacy voucher in Treatment 2.

| | **Sample 1** | **Sample 2** | **Difference** |
|---|---|---|---|
| Sample proportion | 0.2521008 | 0.1 | 0.1521008 |
| 95% CI (asymptotic) | 0.1741 - 0.3301 | 0.0463 - 0.1537 | 0.0556 - 0.2486 |
| z-value | 3.1 | | |
| P-value | 0.002 | | |

Table 7: result of 2-tailed z-test to compare sample proportions

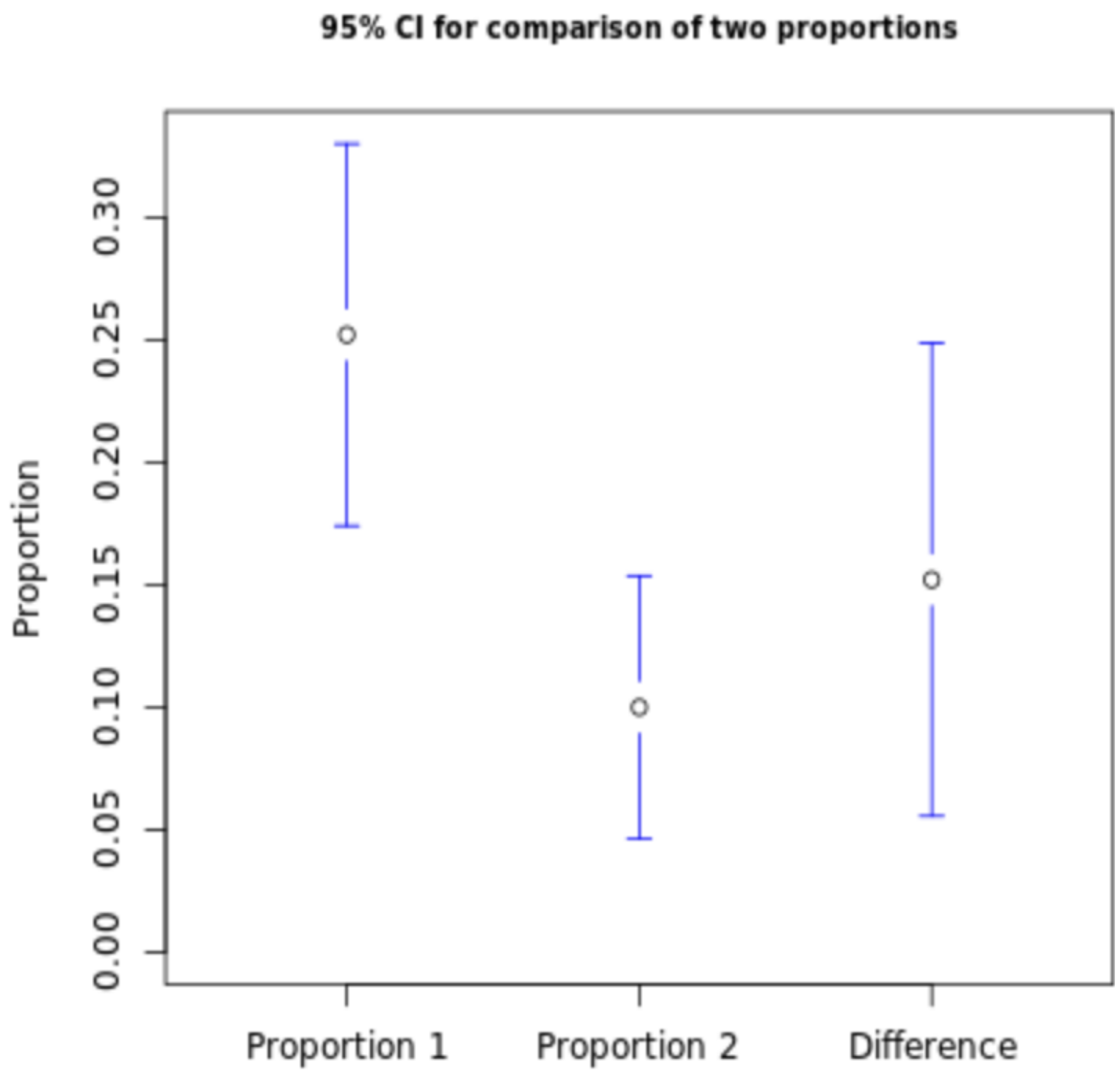

Figure 5: result of 2-tailed z-test to compare sample proportions for Test C

The results of Test C are statistically significant. One should reject the hypothesis that the proportions of participants choosing to pre-commit in the control and in Treatment 2 are equal.